\documentclass[%
 preprint, 
 amsmath,amssymb,
 aps, physrev,
]{revtex4-2}

\usepackage{graphicx}
\usepackage{dcolumn}
\usepackage{bm}

\usepackage[hidelinks]{hyperref}
\usepackage{subcaption}
\usepackage{dirtytalk}

\begin{document}

\title{\textbf{Thermal effects on Buneman Instability: A Vlasov-Poisson Study}
}%
\author{Chingangbam Amudon}
\email[Email: ]{chingangbam.amudon@ipr.res.in}
\affiliation{Institute for Plasma Research (IPR), Gandhinagar, Gujarat, India}
\affiliation{Homi Bhabha National Institute (HBNI), Mumbai, Maharashtra, India}
\author{Sanjeev Kumar Pandey}%
\affiliation{Indian Institute of Technology (IIT) Madras, Chennai, Tamil Nadu, India}
\author{Rajaraman Ganesh}
\email[Corresponding Author: ]{ganesh@ipr.res.in}
\affiliation{Institute for Plasma Research (IPR), Gandhinagar, Gujarat, India}
\affiliation{Homi Bhabha National Institute (HBNI), Mumbai, Maharashtra, India}

\date{\today}
\begin{abstract}
	Buneman Instability has been extensively studied and related aspects namely anomalous resistivity have been explored in detail with the help of analytical theory and  a number of numerical simulations using Particle-In-Cell and Vlasov Solver. The numerical studies have been concentrated in understanding the non-linear evolution of the instability.  In the present study, the growth rate of the Buneman instability in the presence of thermal effects of the constituent species (i.e., ion and electron) is investigated.
	It is observed that the growth rate differs greatly from the one obtained using fluid model (both cold and warm) as well as from the linearized kinetic model. While the well known result of $(m/M)^{1/3}$ dependence of maximum growth rate is recovered in the study, it is shown that the maximum growth rate is essentially independent on the temperature ratio of the constituent species. It is numerically shown that the amplitude of ion's density inhomoneity self-consistently controls the transfer of electron beam energy into the bulk plasma temperature. In particular, as one moves from cold to warm plasma limit, the decrease in ion's inhomogeneity amplitude reduces the generation of sidebands and thus the transfer efficiency.   
\end{abstract}

\maketitle
\newpage
\section{Introduction}
Ion acoustic waves (IAWs) are low frequency waves which are heavily Landau-damped\cite{wong1964landau}, thus making them difficult to be observed in laboratory settings. Buneman\cite{buneman1958} demonstrated that unstable IAWs can be generated and sustained by streaming charge particles, for example current-carrying plasma. In the literature, this phenomenon has been identified under various names such as Buneman Instability\cite{hirose1978},  Ion–Electron Two-stream Instability\cite{yoon2010b}, Current Driven Ion Acoustic Instability (CDIAI)\cite{watt2002ion}, to name a few. It is a type of wave-particle resonant streaming instability. Such IAWs have been observed in astrophysical plasmas, magnetopause, etc\cite{omura2003particle}. 
Hirose\cite{hirose1978} determined that the electric field energy saturation depends on the initial kinetic energy $(W_{0})$ and the mass ratio $(m/M)$, together with the $(m/M)^{1/3}$ dependence of the maximum growth rate where $m$ and $M$ is the electron and ion mass respectively. He also showed that the anomalous resistivity associated with the instability scales as $(m/M)^{2/3}$. 
Related non-linear studies have been performed by Ishihara, Hirose and Langdon\cite{ishihara1980nonlinear, ishihara1981nonlinear, hirose1982nonlinear}, where they identified the existence of algebraic growth -- which was shown to depend on the mass ratio, after the initial exponential growth phase. The final saturation of the instability is believed to be caused by electron trapping in non-sinusoidal potential structures and the wave energy of the instability saturates at $\approx 0.1W_{0}$, independent of mass ratio. These authors further argued that electron trapping leads to the thermalization of the electron.
The non-linear study was followed up by Yoon and Umeda with their `Weak Turbulence' theory\cite{yoon2010b}, in which they address the assumption of Dirac delta velocity distribution function of Ishihara's work with Maxwellian velocity distribution for electron species. The averaged velocity of electron obtained from their Vlasov simulation was shown to be consistent with theoretical predictions, although the temporal evolution of the wave energy shows noticeable deviations.
The effect of collision on the instability has been studied with Particle-In-Cell (PIC) simulation, in which density steepening have been observed at late times\cite{niknam2011simulation, niknam2014particle, shokri2005nonlinear, hashemzadeh2016ion}.

In the past numerical studies which are mostly performed using PIC simulation, the initialization of the particles is such that it corresponds to a Dirac delta distribution function i.e., $f(v) = f_{0}\delta(v-v_{0})$ putting the system in the cold plasma limit. In such studies, we find that the growth rate matches with that of the cold plasma fluid dispersion relation (refer to Section \ref{dispersion}). To the best of our knowledge, the effects of the thermal spread of the distribution function on the growth rate, quasi-linear and non-linear stages of the instability have not been studied in the numerous simulation studies cited above. In the present study, we discuss the effect of thermal spread on the linear, quasilinear and nonlinear dynamics of the Buneman instability using a 1D1V Vlasov-Poisson system. The study is organised as follows -- Section \ref{dispersion} introduce the dispersion relations of the instability and how they are solved, Section \ref{simulation} introduce the newly developed MPI Vlasov-Poisson Solver VPPM-MPI 1.0, Section \ref{result} present the major finding of the study and lastly Section \ref{conclusion} provide the conclusions and perspectives for future work.

\section{\label{dispersion}Analytical Model: Dispersion Relation}
The linearized cold plasma fluid dispersion relation given by Buneman\cite{buneman1958} is as follows,
\begin{equation}
	1 - \frac{\omega_{pe}^{2}}{(\omega - ku_{0})^{2}} - \frac{\omega_{pi}^{2}}{\omega^{2}} = 0
	\label{CPDR}
\end{equation}
which had been solved numerically in the past\cite{buneman1959, hirose1978, rajawat2017, yoon2010b}. When the thermal effects of electron fluid and ion fluid are taken into consideration, the dispersion relation is modified as follows,
\begin{equation}
	1 - \frac{\omega_{pi}^{2}}{\omega^{2}\displaystyle\bigg[1 - \frac{k^{2}}{\omega^{2}} \frac{\gamma_{i} k_{B}T_{i}}{m_{i}}\bigg]} - \frac{\omega_{pe}^{2}}{(\omega - u_{0}k)^{2}\displaystyle\bigg[1 - \frac{ k^{2}}{(\omega - u_{0}k)^{2}}\frac{\gamma_{e} k_{B}T_{e}}{m_{e}}\bigg]} = 0
	\label{WPDR}
\end{equation}
where $\gamma_{i}$ and $\gamma_{e}$ are the adiabatic constant for ion and electron respectively. Following standard linearization procedure, a kinetic dispersion relation is derived from the coupled Vlasov-Ampère equation\cite{raghunathan2013nonlinear}. The linearized kinetic dispersion relation is as follows,
\begin{equation}
	\frac{4\pi e^{2}}{m_{i}}\int v_{i}\left(\frac{\partial_{v}f_{i0}}{kv_{i}-\omega}\right)dv + \frac{4\pi e^{2}}{m_{e}}\int v_{e} \left(\frac{\partial_{v}f_{e0}}{kv_{e}-\omega}\right)dv-\omega =0
	\label{LKDR}
\end{equation}
The above equation (Eqn. \ref{LKDR}) represents a dispersion relation, wherein both electrons and ions are treated kinetic. Eqn. \ref{LKDR} may be reduced in terms of Plasma Dispersion function by considering both $f_{i0}$ and $f_{e0}$ to be Maxwellian resulting in the following equation,
\begin{equation}
	1- \frac{1}{2k^{2}}Z'(\xi_{e})-\frac{1}{2T_{r}k^{2}}Z'(\xi_{i})=0
	\label{PDFR}
\end{equation}
where the streaming information is included in the definition of $\xi_{e}$ and $\xi_{i}$ for both species, $M_{r} = m_{i}/m_{e}$ is the mass ratio and $T_{r} = T_{i}/T_{e}$ is the temperature ratio.
\begin{equation}
	\xi_{e} = \displaystyle\frac{\omega/k-u_{e0}}{\sqrt{2} v_{th, e}};\hspace{2em}\xi_{i} = \displaystyle\frac{\omega/k-u_{i0}}{v_{th, e}}\sqrt{\frac{M_{r}}{2T_{r}}}
\end{equation}
where $v_{th,e}$ is the thermal velocity of electron and $u_{e0}/u_{i0}$ is the streaming velocity of electron/ion. All the above dispersion relations are normalized to electron parameters i.e., length $(x)$ by Debye length $(\lambda_{De})$, time $(t)$ by plasma frequency $(\omega_{pe})$ and velocity $(v)$ by electron thermal velocity $(v_{th,e})$  for consistency and solved numerically. In Figure \ref{growth_rate}, a comparison of dispersion relations is shown, which is useful for validating the results of numerical simulations and for identifying physics effects. The roots of the fluid dispersion relations, i.e., Eqns.\ref{CPDR} and \ref{WPDR}, are computed by Newton-Raphson method after decomposing them into their real and imaginary components by assuming a complex frequency $\omega = \omega_{r} + i\omega_{i}$.  Similar decomposition is applied to the kinetic case i.e., Eqn. \ref{LKDR} and the roots are extracted by locating the zero-crossing points of the real and imaginary components. In the next section, the mathematical model of the nonlinear 1D1V Vlasov-Poisson system used in the simulation is presented.

\section{\label{simulation}Mathematical Model: 1D1V Vlasov-Poisson Solver}
The numerical simulation is performed by an in-house developed MPI Vlasov-Poisson Solver, VPPM-MPI 1.0 which is the upgraded version of VPPM-OMP 1.0, an OpenMP code\cite{sanjeev, pallavi}. The mathematical model consists of the 1D1V coupled Vlasov–Poisson equations,
\begin{subequations}
	\begin{equation}
		\frac{\partial f_{e}}{\partial t}+  v_{e}\cdot \frac{\partial f_{e}}{\partial x} - E\cdot\frac{\partial f_{e}}{\partial v_{e}} = 0 
	\end{equation}
	\begin{equation}
		\frac{\partial f_{i}}{\partial t}+v_{i}\cdot\frac{\partial f_{i}}{\partial x}+\frac{E}{M_{r}}\cdot\frac{\partial f_{i}}{\partial v_{i}}=0
	\end{equation}
	\begin{equation}
		\frac{\partial E}{\partial x} = \int f_{i} dv_{i}  - \int f_{e}dv_{e}
	\end{equation}
\end{subequations}
where $f_{e}(x, v_{e}, t)$ and $f_{i}(x, v_{i}, t)$ is the normalized electron and ion distribution function respectively, $E(x, t)$ is the self consistent normalized electric field and $M_{r}$ is the mass ratio. The above equations have been normalized to electron parameters (refer Section \ref{dispersion}).

The Vlasov equations are reduced to a set of coupled advection equation with Time Splitting Scheme\cite{cheng1976}. Piecewise Parabolic Method (PPM)\cite{collela1984} is used to solve each of the advection equation.
Simulation domain in phase space is chosen to be $(0, L_{max})$ in space and $(-v_{s}^{max}, v_{s}^{max})$ in velocity, where $L_{max} = 2\pi/k_{min}$ is the system size, $s = i, e$ are the species and $v_{e}^{max}$  is the maximum electron velocity. Periodic boundary conditions (PBC) have been implemented in both spatial and velocity domains and the domain is discretized into $(N_{x} \times N_{v})$ grids. Adaptive time step has been implemented with $dt_{max} = 0.1$ for maintaining the details as well as the accuracy of the time evolution. With the numerical methodology for solving the Vlasov–Poisson system established, the subsequent section presents the results of the simulations where the thermal effects on linear and nonlinear phases of Buneman instability are elucidated.

\section{\label{result}Results and Discussions}
The simulation is setup initially with a stationary ion species, acting as a neutralizing background alongside a streaming electron species with spatial perturbation, $g(x)$; both species are initialized with Maxwellian velocity distribution. Mathematically,
\begin{subequations}
	\begin{equation}
		f(x, v_{e}, 0)  = \left(1+g(x))\right)\cdot \frac{1}{\sqrt{2\pi}}\exp\left(-\frac{(v_{e}-u_{0})^{2}}{2}\right)
		\label{init_electron}
	\end{equation}
	\begin{equation}
		f_{i}(x, v_{i}, 0) = \frac{1}{\sqrt{2\pi}}\sqrt{\frac{M_{r}}{T_{r}}}\cdot \exp\left(-\frac{{v}_{i}^{2}}{2}\cdot\frac{M_{r}}{T_{r}}\right)
		\label{init_ion}
	\end{equation}
	\text{and}
	\begin{equation}
		g(x) = \begin{cases}
			\alpha\cdot\sin(k_{perp}\cdot x + \phi) &; -\pi \leq \phi \leq \pi \; (\text{Fourier}) \\
			\alpha\cdot r &; - 1 \leq r \leq 1 \; (\text{White Noise})
		\end{cases}
		\label{pert_mode}
	\end{equation}
\end{subequations}
where $\alpha = 0.01$ is the amplitude of perturbation, $M_{r} = 1836$, $T_{r}=0.1$, $k_{perp}$ is the wave number in which the perturbation is given ($k_{perp} = nk_{min}$ where $ n = 1, 2, 3\dots$), $\phi$ is a random phase, $r$ is random number and $u_{0}$ is the streaming/drifting velocity of electron. In this work, two types of perturbation are implemented -- (a) Fourier perturbation and (b) White Noise perturbation as shown in Eqn. \ref{pert_mode}. The values of amplitude of perturbation, mass ratio and temperature ratio remain the same throughout unless otherwise stated. All the parameters given in this section are in normalized units.

\begin{figure}[ht]
	\centering
	\begin{subfigure}{0.49\textwidth}
		\includegraphics[width=\linewidth]{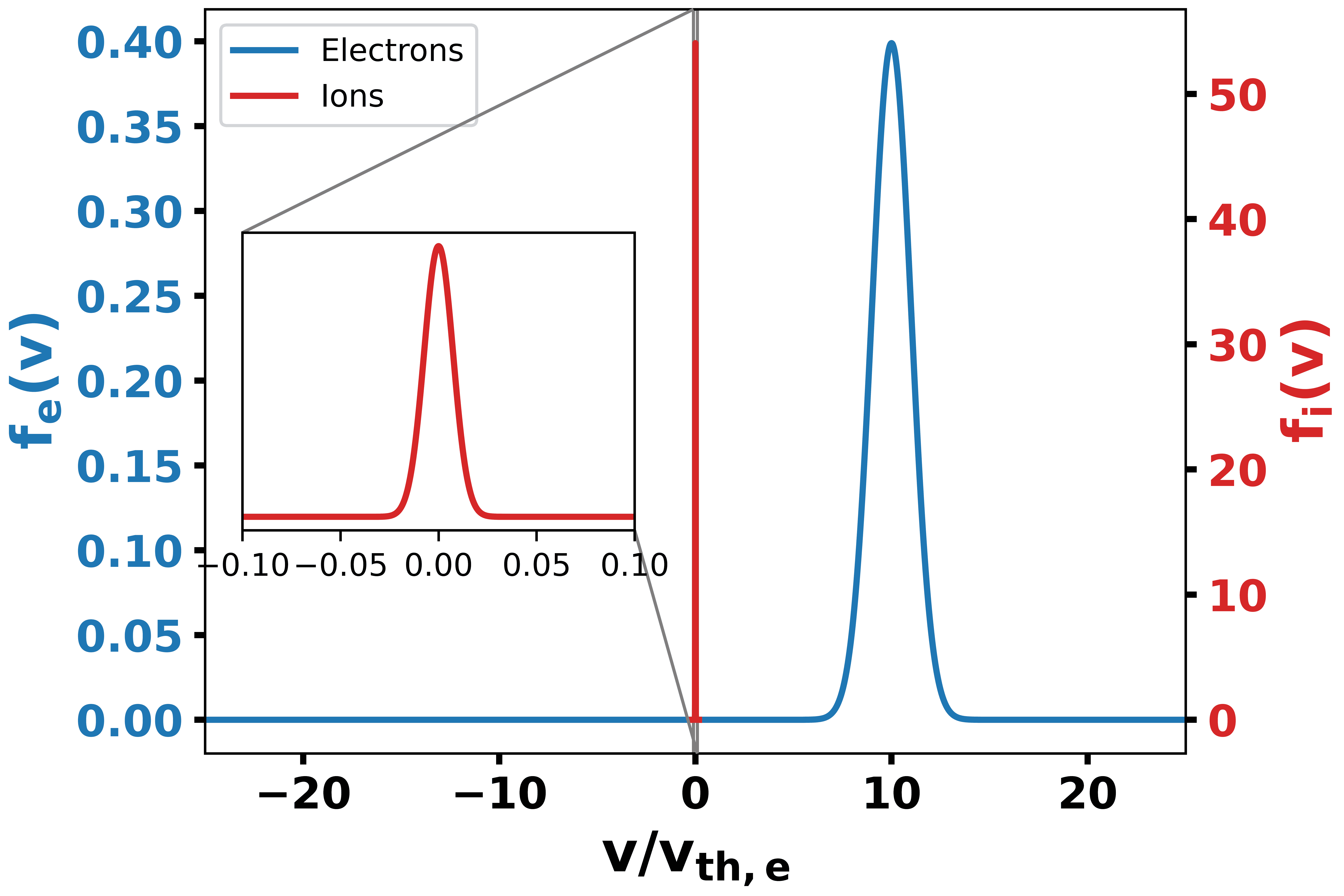}
		\caption{}
		\label{cold_average}
	\end{subfigure}
	\begin{subfigure}{0.49\textwidth}
		\includegraphics[width=\linewidth]{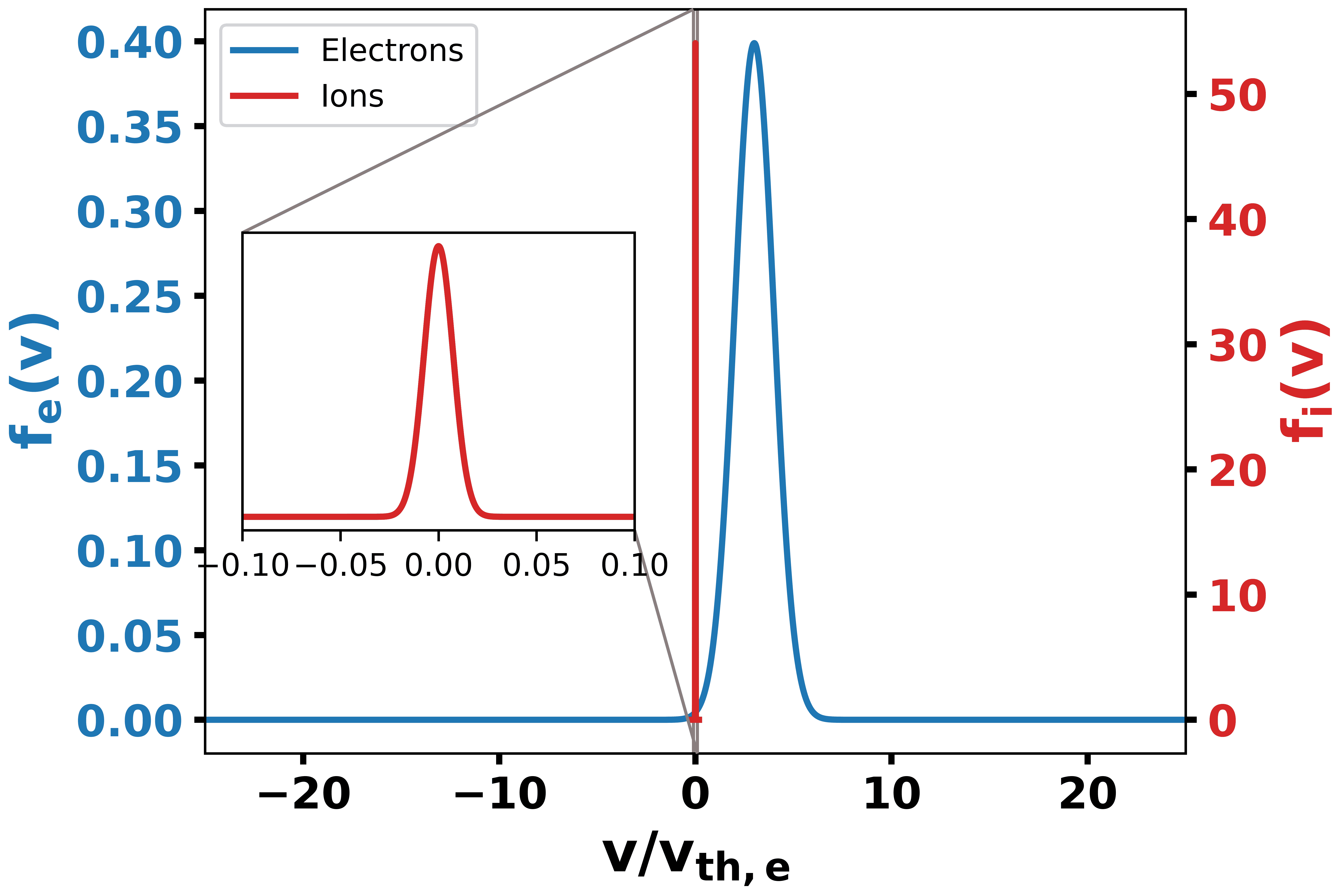}
		\caption{}
		\label{warm_average}
	\end{subfigure}
	\caption{Space-averaged Electron and Ion Velocity Distribution i.e., $\hat{f}_{e}(v)\; \& \; \hat{f}_{i}(v)$ at (a) Cold Plasma Limit $(u_{0} = 10)$ and (b) Warm Plasma Limit $(u_{0} = 3)$ where the streaming velocity is normalized to electron thermal velocity ($v_{th, e}$).}
	\label{space-averaged_f}
\end{figure}
One of the difficulties of Vlasov simulation is reaching the cold plasma limit -- sampling of a Dirac delta function can be achieved in a Eulerian grid when $dv\to 0$ i.e, very high resolution. Such sampling increases the computational cost exponentially without much improvement in resolution. Another approach is taken to reach the cold plasma limit where the electron velocity distribution is shifted such that the thermal part does not overlap with the ion velocity distribution function. Thermal effects will be observed as there is finite width of the electron velocity distribution.
The different plasma limit is shown in Figure \ref{space-averaged_f}, the cold plasma limit is characterized with the ion distribution far from the thermal part of the electron distribution while in the warm plasma limit, the ion falls near or under the thermal part of the electron distribution function. 
The parameters are set in such a way that the system is in the cold plasma limit -- long wavelength and high streaming velocity i.e., $k_{perp} = 0.1 (=k_{min})$ with streaming velocity range, $6.0 \leq u_{0} \leq 11.0$. The grid resolution of the cold plasma simulation is $(N_{x}, N_{v}) = (2048, 16384)$ with $v_{e}^{max} = 25$.

\begin{figure}[ht]
	\centering
	\includegraphics[width=0.8\linewidth]{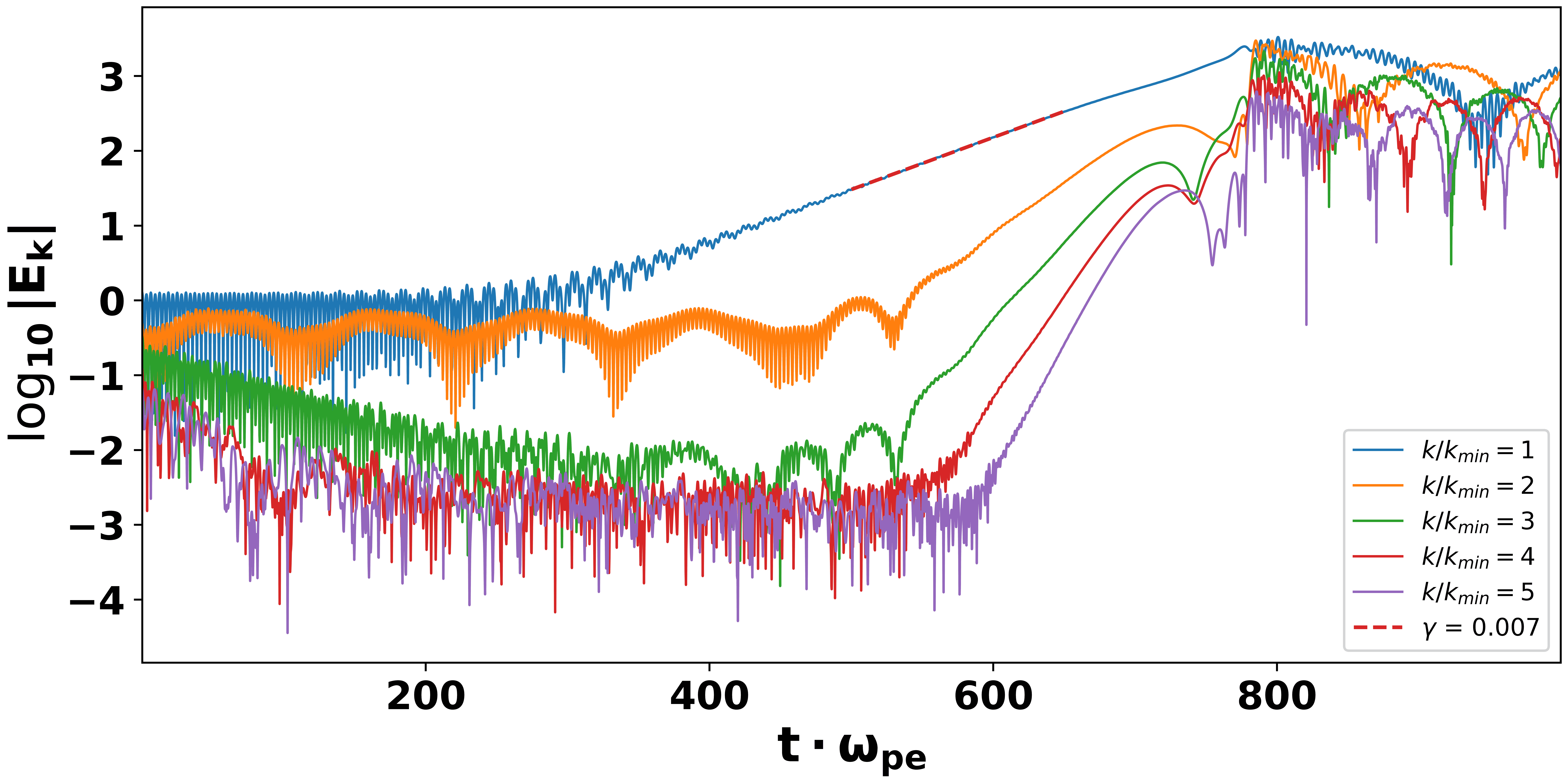}
	\caption{Temporal Evolution of Fourier Electric mode (fundamental and higher harmonics) for $k_{min}=0.1$ and $u_{0} = 6.0$. The perturbation is given to wave number $k_{perp} = 0.1$ or $k_{perp}/k_{min} = 1$ while $k/k_{min} = 2, 3, ...$ are generated self consistently. The frequency of the fundamental mode is $1.0156$, corresponds to Langmuir frequency and it is observed that different modes are of comparable amplitude after saturation of the instability.}
	\label{electric_mode_cold}
\end{figure}
The temporal evolution of Fourier electric modes -- fundamental and higher harmonics is shown in Figure \ref{electric_mode_cold}. The wave number of the initial Fourier perturbation is $k_{perp} = k_{min} = 0.1$ and the higher modes (i.e., $k/k_{min} = 2, 3\dots$) are excited self consistently due to nonlinear dynamics as the system evolves. It is observed that the fundamental mode i.e., $k_{min} = 0.1$ is the dominant mode while the higher harmonics grow at later times. The frequency associated with the fundamental mode is $\omega = 1.0156$, which corresponds to the Langmuir frequency of the considered mode. The wave-particle interaction leads to the injection of energy from the streaming electron to the wave, which in turns excites the IAWs. It is observed that only for a certain range of streaming velocity, the wave-particle interaction occurs thus confirming that the instability is a resonant streaming instability. After saturation of the instability, the modes become comparable in amplitude and the mode-mode coupling allows higher harmonics to become briefly dominant before the fundamental mode again overtakes them.

\begin{figure}[ht]
	\centering
	\includegraphics[width=\textwidth]{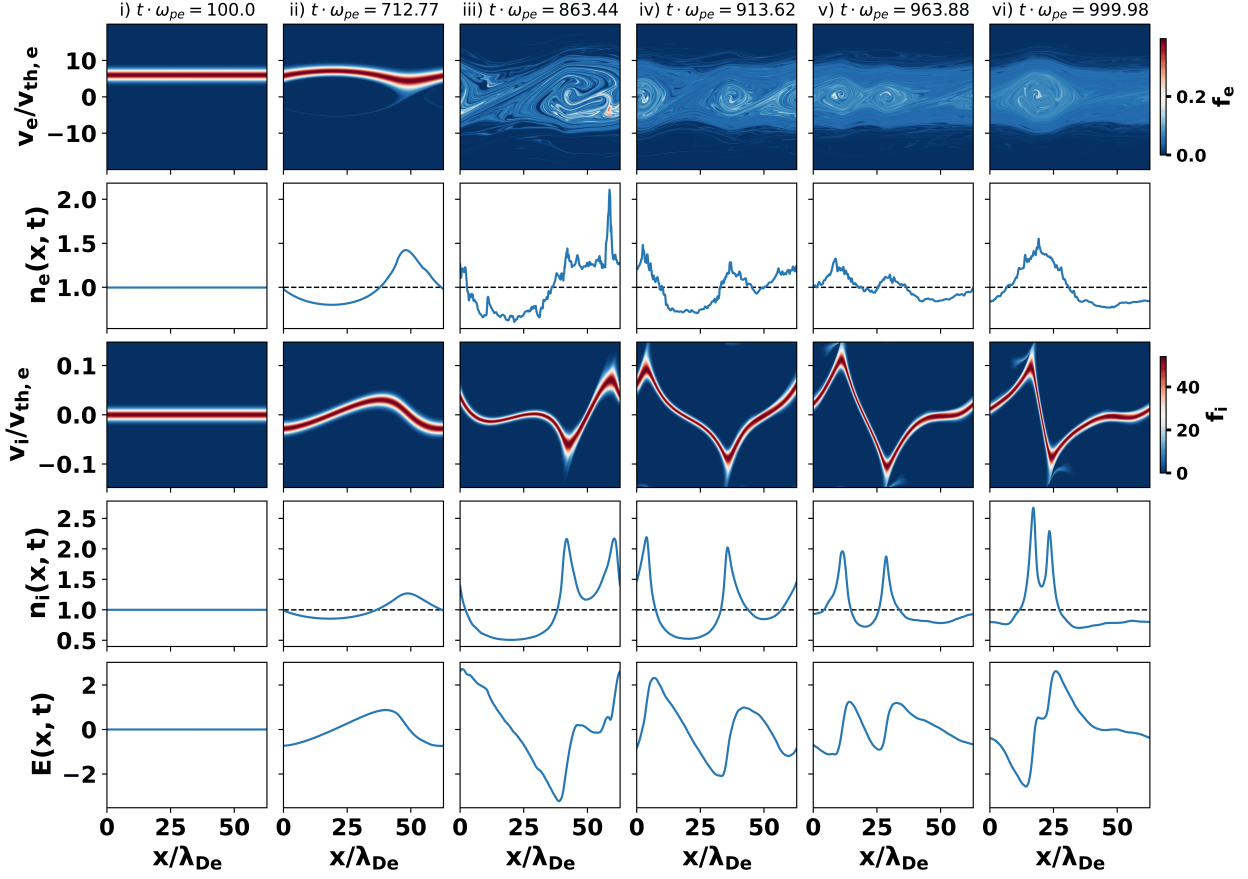}
	\caption{Phase Space Information of electron (first row) and ion (third row) with their density profile (second and fourth row respectively) and electric field (fifth row) at different time of simulation. Density steepening is observed for both species.}
	\label{cold_phase}
\end{figure}
Figure \ref{cold_phase} shows the electron phase space (first row), electron density (second row), ion phase space (third row), ion density (fourth row), and electric field (fifth row) at different times during the simulation. The change in the dominant mode leads to the tearing of a large electron phase-space hole into smaller structures, as illustrated in Figure \ref{cold_phase} (first row - column iii and iv). Such tearing has previously been reported by Rajawat \textit{et al.}\cite{rajawat2017} using PIC simulations.
Additionally in their work, the well-known phenomenon of density steepening for both species is observed -- electron and ion densities steepen to nearly or more than ten times the initial density. 
In contrast, the steepening in the present simulation reaches only about twice the initial density as shown in the second and fourth rows of Figure \ref{cold_phase}. This reduced steepening is due to the finite thermal spread of the considered species.
In the present study, the electron hole-ion soliton coupling is captured with high precision and resolution (see the electron phase space, first row and ion density, fourth row in column iv and v of Figure \ref{cold_phase}). Furthermore, the electron's streaming/beam energy is entirely transferred into the instability and subsequent electron thermalization. As a result, the beam component of the electron distribution vanishes, leaving behind a broadened thermal component with embedded phase-space structures.

\begin{figure}[ht]
	\centering
	\includegraphics[width=0.85\linewidth]{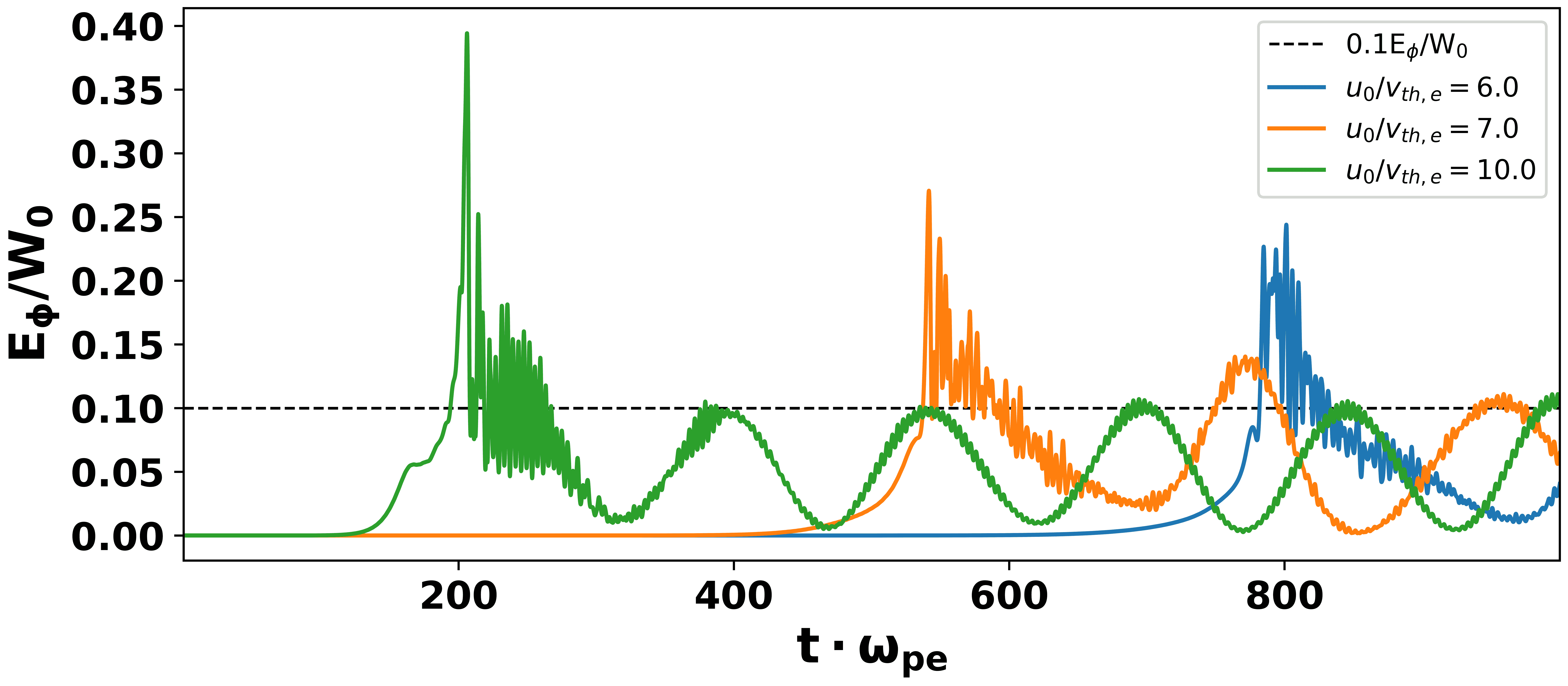}
	\caption{Temporal Evolution of the Ratio of Potential Energy with initial Kinetic Energy, $\text{E}_{\phi}/\text{W}_{0}$. A surge in potential energy is observed at the time of saturation, then reduces to $\approx 0.1\text{W}_{0}$.}
	\label{cold_electrostatic_energy}
\end{figure}

Another important characteristic of the instability in cold plasma limit is the bound on the electrostatic energy approximately to $0.1\text{W}_{0}$ (where $\text{W}_{0}$ is the initial kinetic energy of electron), which was first identified by Ishihara\cite{ishihara1980nonlinear}. In the present work, the bound is shown graphically by the temporal evolution of the ratio of electrostatic energy with initial kinetic energy i.e., $E_{\phi}/\text{W}_{0}$ in Figure \ref{cold_electrostatic_energy} for various values of streaming velocity of electron in which the energy increases during the linear stage of instability, then a sudden spike at the time of saturation and after saturation it exhibits oscillatory behavior with $\approx0.1\text{W}_{0}$ as the upper bound. The sudden spike in potential energy at the time of saturation is caused by the abrupt electron trapping leading to the injection of all the streaming energy into electrostatic energy, followed by the thermalization of electron.

\begin{figure}[ht]
	\centering
	\includegraphics[width=0.6\textwidth]{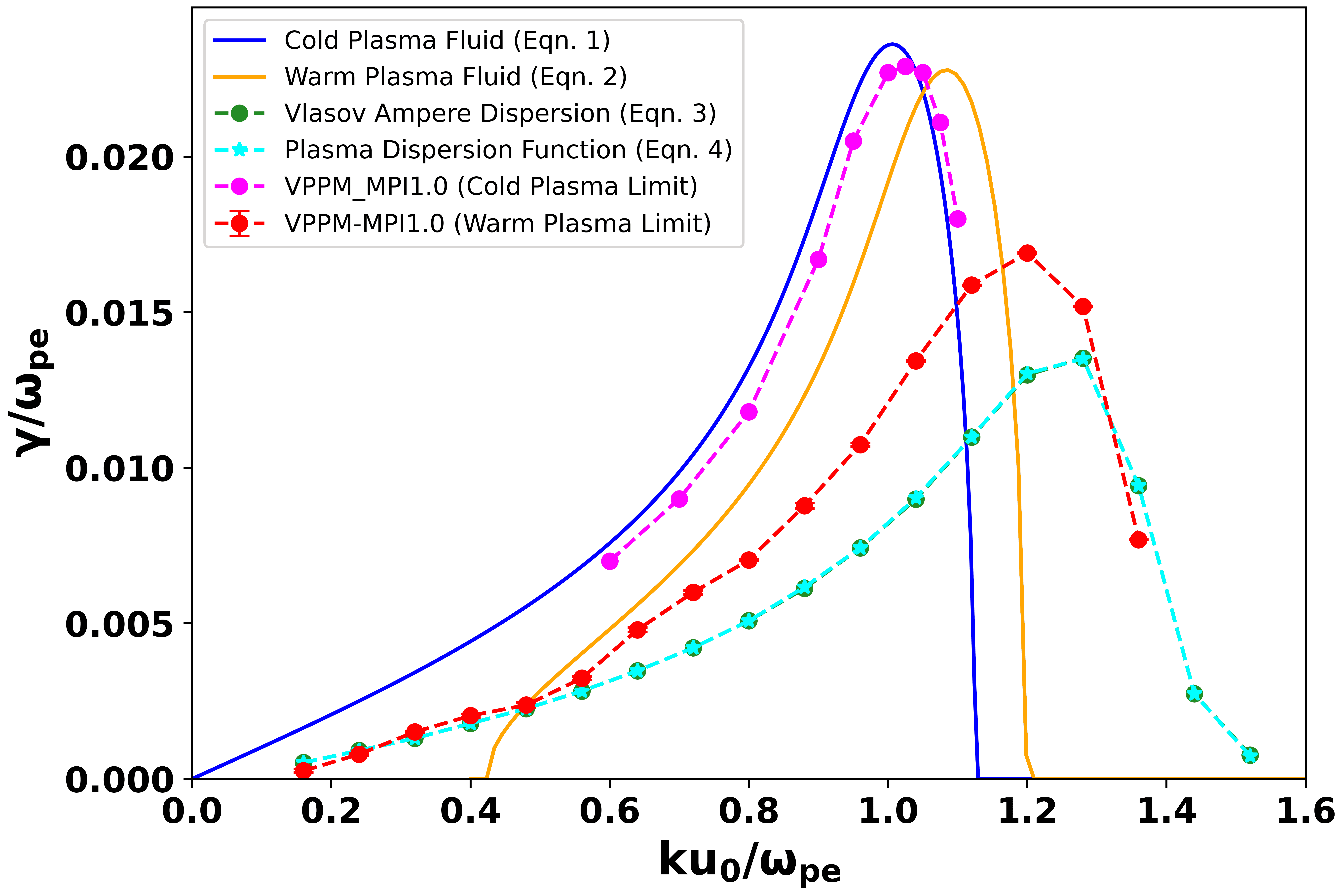}
	\caption{Growth Rate (in Base 10) of Buneman instability for the different plasma limit -- magenta for cold plasma limit while red for warm plasma limit. The simulation growth rate at warm plasma limit (red) is slightly higher than the kinetic model, but lower than fluid model.}
	\label{growth_rate}
\end{figure}

Akin to the PIC simulation results in the cold plasma limit\cite{ishihara1980nonlinear, shokri2005nonlinear, niknam2011simulation, niknam2014particle, rajawat2017} -- the phenomenon of density steepening, the surge in electrostatic energy -- have been reproduced successfully with the Vlasov Solver. The growth rate at the said limit (magenta) is plotted and benchmarked against the cold plasma fluid dispersion relation (Eqn. \ref{CPDR}) in Figure \ref{growth_rate}, where the well-known results of cold plasma fluid dispersion relation -- (a) the growth rate depends only on Doppler shift, $ku_{0}$ i.e., any values of $k$ and $u_{0}$ having the same value of $ku_{0}$ will have the same growth rate and (b) the maximum growth rate occurs at $ku_{0} = 1$ -- are observed. The simulation results qualitatively match the analytical results. The growth rate are defined in base 10 values. The deviation seen in the Figure \ref{growth_rate} between the cold plasma simulation and the cold plasma fluid dispersion relation is associated to the thermal effects as Vlasov simulation can not resolve a true Dirac delta distribution function. With the cold plasma results established, the behavior of the instability in the warm plasma limit is examined in the following.

\begin{figure}[ht]
	\begin{subfigure}{0.59\textwidth}
		\centering
		\includegraphics[width=\linewidth]{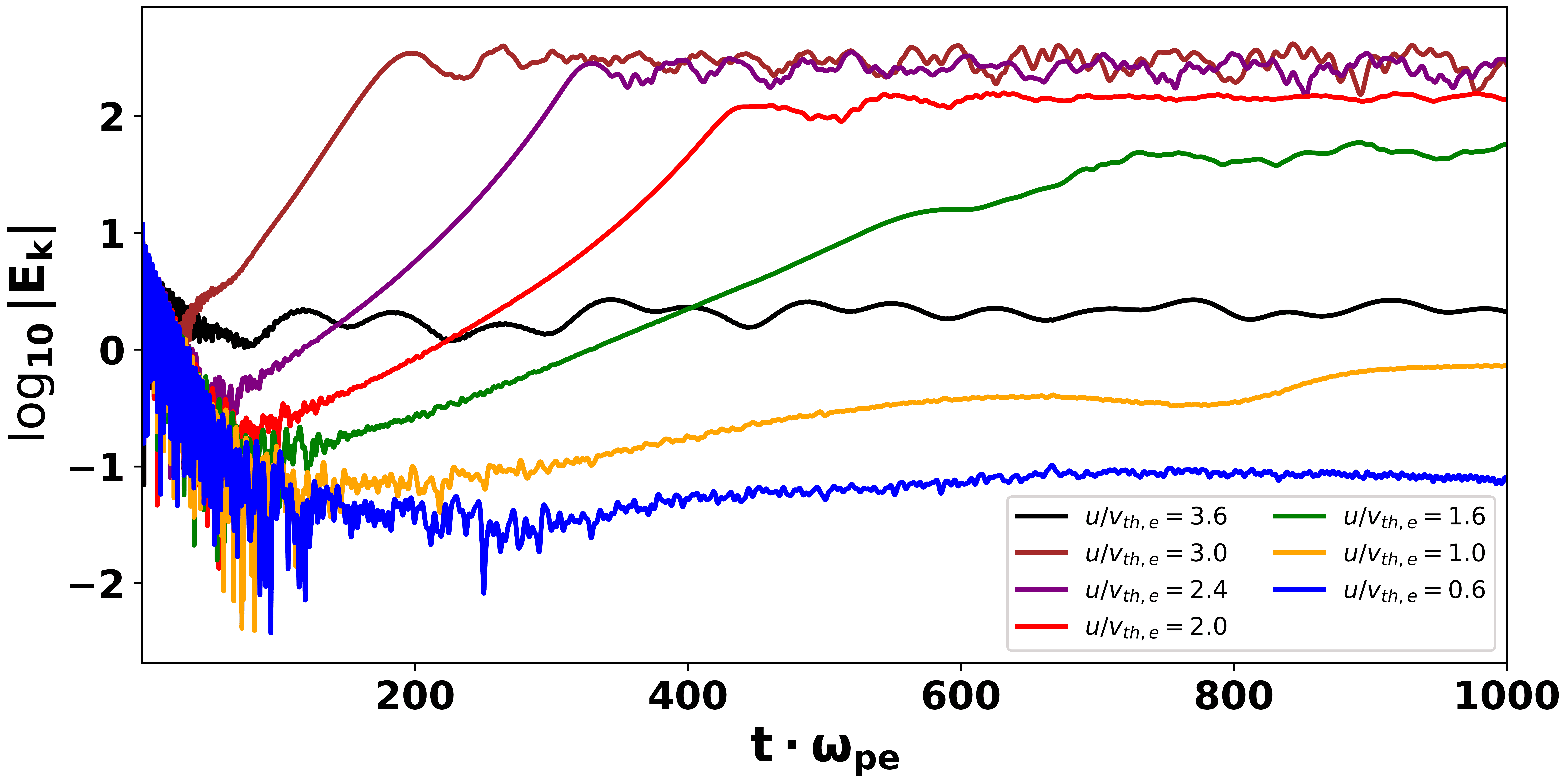}
		\caption{}
		\label{elec_fourier}
	\end{subfigure}
	\begin{subfigure}{0.392\textwidth}
		\centering
		\includegraphics[width=\linewidth]{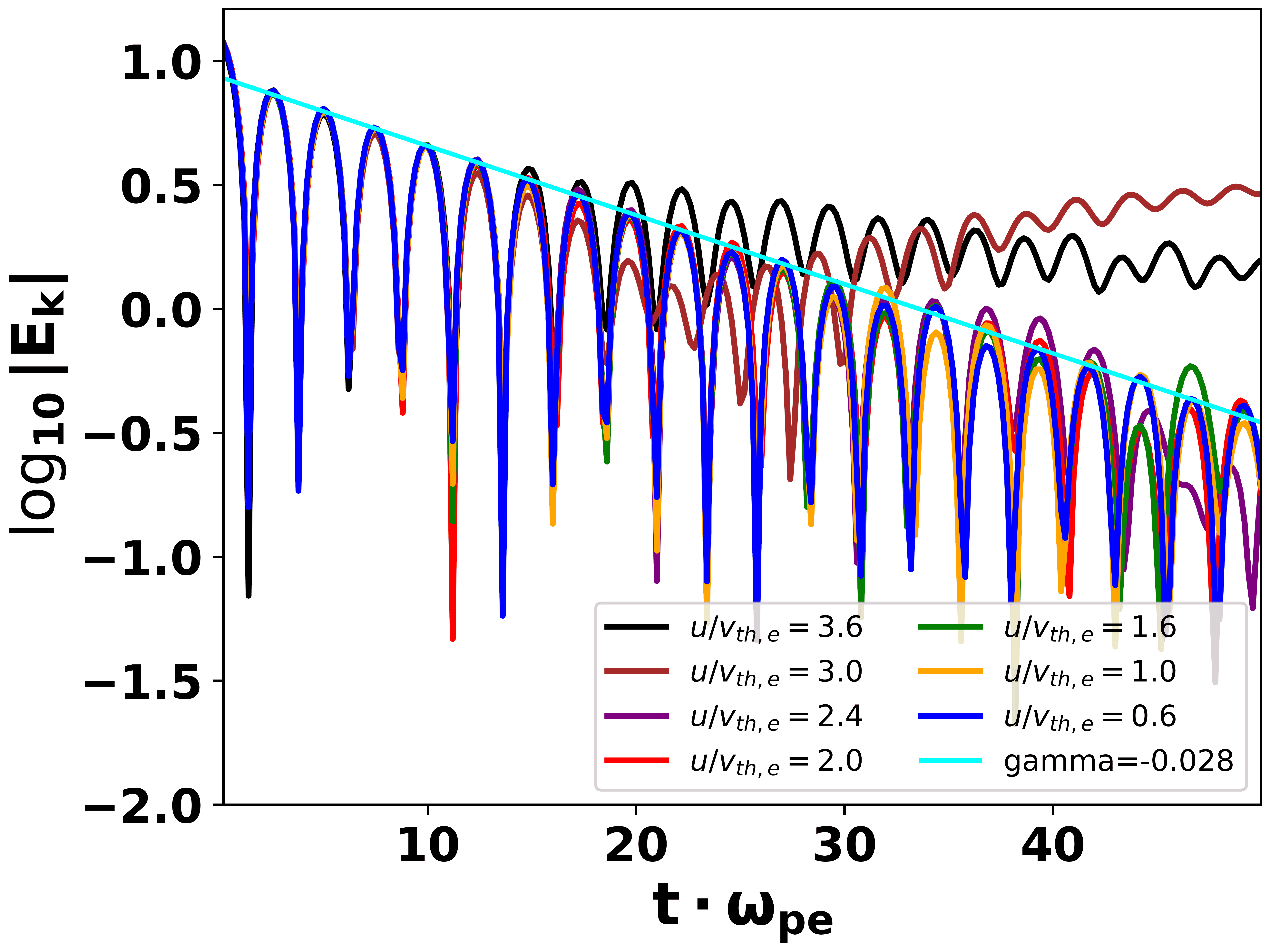}
		\caption{}
		\label{elec_damping}
	\end{subfigure}
	\caption{Temporal Evolution of dominant Fourier Electric mode, $E_{k_{min}=0.4}$ for various value of streaming velocity, $u_{0}$ from (a) $t\cdot\omega_{pe} = 0\;\text{to}\;1000$ and (b) $t\cdot\omega_{pe} = 0\;\text{to}\;50$. From Fig. \ref{elec_fourier}, one observes that the wave-particle interaction occurs for only a small range of streaming velocity with $u_{0} = 3.0$ having the maximum growth rate. From Fig. \ref{elec_damping}, Landau damping of the initial perturbation for all values of streaming velocity is observed with $\omega = 1.281$ and $\gamma = -0.028$.}
	\label{elec_fourier_pert}
\end{figure}

The electron streaming velocity is reduced such that ion lies near the thermal part of electron as shown in Figure \ref{warm_average} i.e., warm plasma limit. By varying the value of streaming velocity of electron, $0.2 \leq u_{0} \leq 3.8$ and fixing the wave number at $k_perp = k_{min} = 0.4$, the instability is studied for a range of Doppler shift values. The grid resolution for the simulation is $(N_{x}, N_{v}) = (1024, 8192)$ with $v_{e}^{max} = 12.0$. The temporal evolution of dominant Fourier electric mode, $E_{k_{min} = 0.4}$ for various value of streaming velocity is given in Figure \ref{elec_fourier_pert}. At the beginning of the simulation, Landau damping of the perturbation is observed over the whole range of streaming velocity as shown in Figure \ref{elec_damping}. From the Figure \ref{elec_damping}, the frequency of oscillation, $\omega = 1.281$ and the damping rate, $\gamma = -0.028$ are determined which is consistent with the Langmuir frequency for the corresponding mode.

The growth rate (red in color) of the instability determined from the simulation with Fourier perturbation for various values of Doppler shift, $ku_{0}$ or streaming velocity, $u_{0}$, is shown in Figure \ref{growth_rate}. Four runs are performed with random phases for a particular set of parameter to ensure the growth rates are not numerical artifacts. Owing to the absence of noise in the Vlasov-Poisson simulation, the growth rate for a given set of parameters remains unchanged, resulting in negligible error bars. Fluid (Eqn. \ref{WPDR}) and kinetic (Eqn. \ref{LKDR} and Eqn. \ref{PDFR}) dispersion relations are used to validate the results of the simulation.

From Figure \ref{growth_rate}, it is observed that the introduction of thermal effects into the mathematical model introduces correction on the growth rate of the instability (refer Appendix \ref{decoupling} for details), thus deviates from the cold plasma fluid results; the maximum growth rate has decreased and the corresponding Doppler shift has shifted to higher value for warm plasma fluid dispersion relation. The Plasma Dispersion function relation (Eqn. \ref{PDFR}) is solved with the standard Python library, \textbf{PlasmaPy}\cite{plasmapy}. In Figure \ref{growth_rate}, the numerical solutions of the kinetic dispersion relations overlap one another indicating the accuracy of the developed Dispersion Solver (Eqn. \ref{LKDR}).  The fluid dispersion relation (Eqn. \ref{WPDR}) overestimates the growth rate while the kinetic dispersion relations (Eqn. \ref{LKDR} and Eqn. \ref{PDFR}) are able to match the smaller values of streaming velocity $u_{0}$ (i.e., $ku_{0}< 0.5$), but fails at higher values for the warm plasma simulation. 

\begin{figure}[ht]
	\centering
	\begin{subfigure}{0.48\textwidth}
		\includegraphics[width=\linewidth]{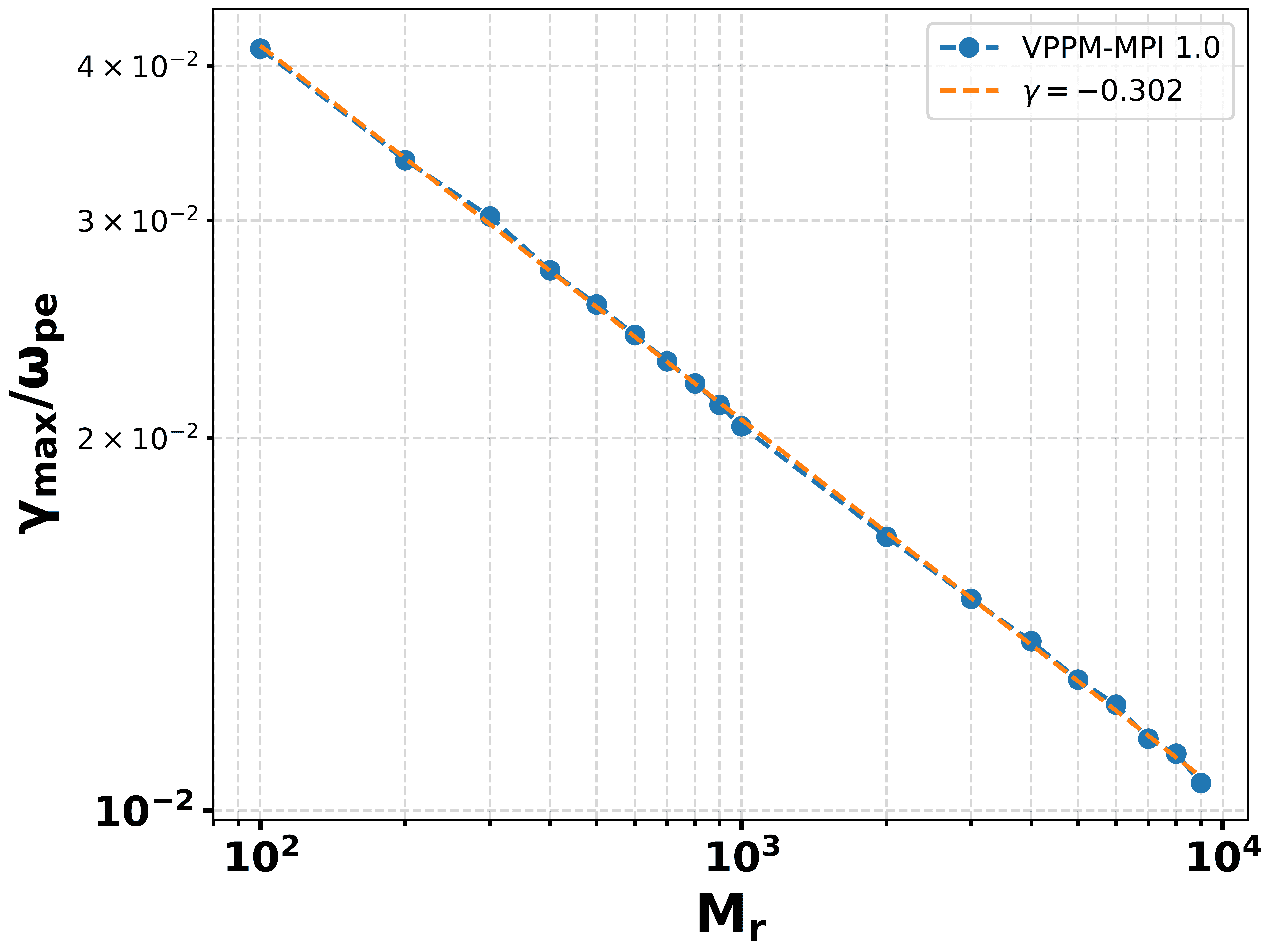}
		\caption{}
		\label{mr_gamma}
	\end{subfigure}
	\begin{subfigure}{0.48\textwidth}
		\includegraphics[width=\linewidth]{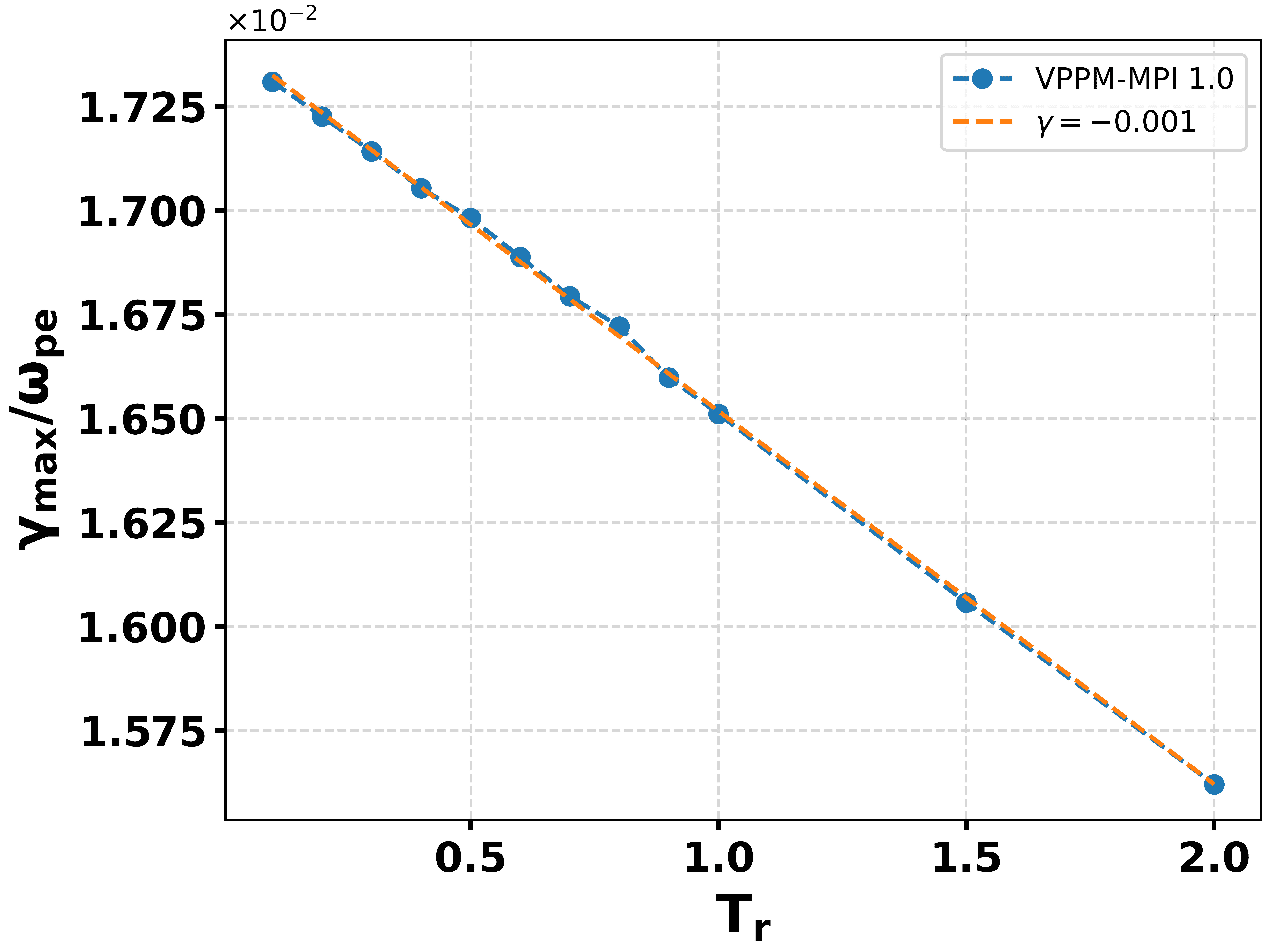}
		\caption{}
		\label{tr_gamma}
	\end{subfigure}
	\caption{Variation of Maximum Growth Rate, $\gamma_{max}$ (i.e., $k  = 0.4, u_{0} = 3.0$) w.r.t. (a) Mass Ratio $M_{r}$ for $T_{r} = 0.1$ and (b) Temperature Ratio, $T_{r}$ for $M_{r} = 1836$. One observes the established $M_{r}^{-1/3}$ dependence in Fig. \ref{mr_gamma}; the growth rate is  nearly independent of Temperature Ratio, $T_{r}$ in Fig. \ref{tr_gamma}.}
\end{figure}

The maximum growth rate of the Buneman instability has been shown analytically to depend on the reciprocal of the cube root of mass ratio i.e. $\gamma_{max} \propto M_{r}^{-1/3}$ as discussed in various refs\cite{buneman1958, buneman1959, hirose1978, ishihara1980nonlinear}. In the present study, this dependence has been investigated over a range of mass ratio, $100 \leq M_{r}  \leq 9000$ for parameters corresponding to the value of Doppler shift where growth rate is maximum i.e., $k=0.4, u=3.0$ and $T_{r} = 0.1$.  Figure \ref{mr_gamma} presents the plot between maximum growth rate and mass ratio. A linear fit to the data yields a slope of $-0.302$, which is in close agreement with the theoretical expectation\cite{niknam2011simulation, rajawat2017}.

The dependence of the maximum growth rate on the temperature ratio is also explored in the present study. A range of temperature ratio, $0.1 \leq T_{r} \leq 2.0$ is considered with $k=0.4$, $u_{0} = 3.0$ and  $M_{r} = 1836$; it can be seen in Figure \ref{tr_gamma} that the maximum growth rate is nearly independent of the temperature ratio. In standard Plasma Physics books\cite{goldston2020introduction, krall1973principles}, the growth rate of the instability is independent of temperature ratio for $T_{r} \ll 1$. The present result generalizes the dependency by showing that the ion contribution to the maximum growth rate is negligible even at $T_{r} \gtrsim 1$. 

\begin{figure}[ht]
	\centering
	\includegraphics[width=0.6\textwidth]{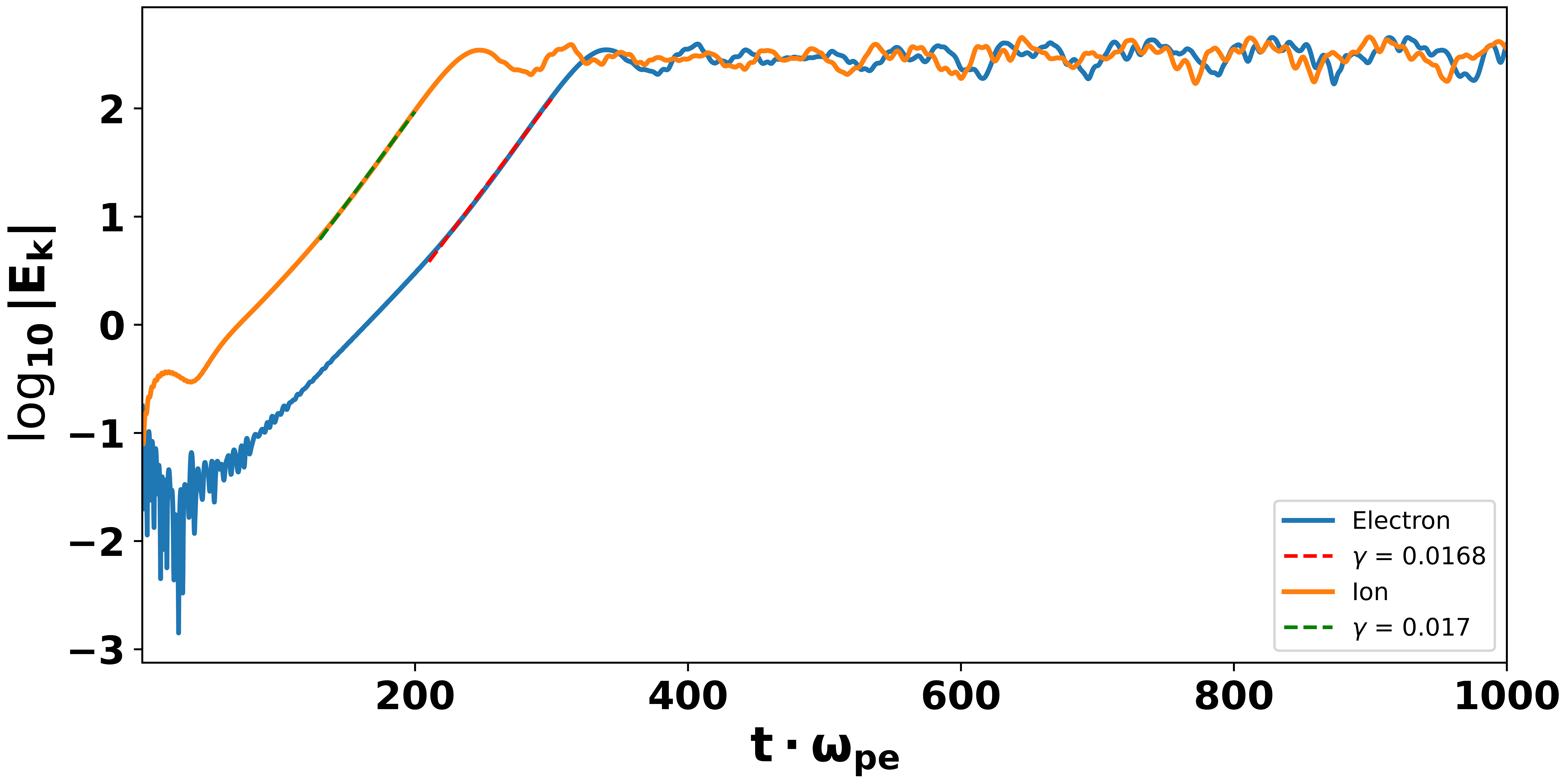}
	\caption{Temporal Evolution of dominant Fourier Electric mode for White Noise perturbation i.e., $g(x) = \alpha \cdot r$ (where $-1\leq r \leq 1$ is a random number). The blue curve represents the perturbation applied to the electron, while the orange corresponds to the ion. In both cases, the growth rate are approximately the same i.e., $\gamma \approx 0.017$.}
	\label{white_noise_fourier_mode}
\end{figure}
The growth rate of the instability should be independent of the perturbed species and mode of perturbation. Simulations have been performed with white noise perturbation i.e., $g(x) = \alpha \cdot r$ (where $-1\leq r \leq 1$ is a random number) rather than Fourier perturbation for streaming velocity, $u_{0} = 3.0$ in which the perturbed species is either electron or ion; Figure \ref{white_noise_fourier_mode} shows the evolution of the amplitude of the dominant Fourier electric mode. The white noise perturbation suppressed the Langmuir wave signature for ion-perturbed simulation, helping in the accurate determination of the growth rate. The growth rate $(\gamma \approx 0.017)$ remains the same for both runs as seen in Figure \ref{white_noise_fourier_mode}, thus confirming that the growth rate of instability remains the same even with different mode of perturbation and perturbed species. The difference is the time of saturation of the instability i.e., ion-perturbed simulation saturates at earlier time as compared to electron-perturbed simulation; Landau damping is observed in the latter, but not in former.

\begin{figure}[ht]
	\centering
	\includegraphics[width=0.8\textwidth]{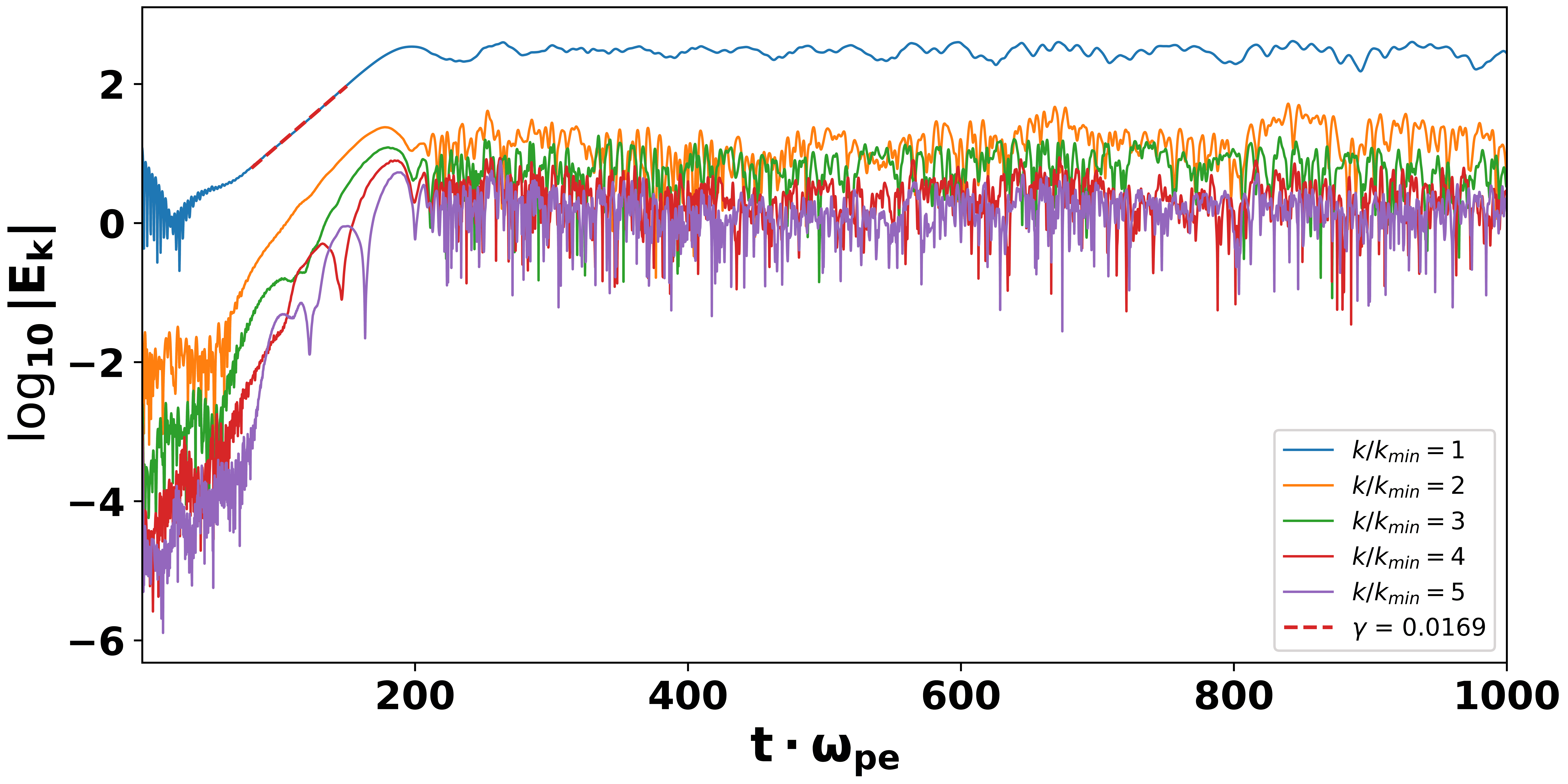}
    \caption{Temporal Evolution of Fourier Electric mode (fundamental and higher harmonics) for $k_{perp}=0.4$ and $u_{0} = 3.0$. The perturbation is given to wave number $k_{perp} = 0.4$ while $k/k_{min} = 2, 3\dots$ are generated self consistently. It is observed that fundamental mode is at least two order higher than its harmonics in amplitude after saturation.}
    \label{electric_mode_warm}
\end{figure}
\begin{figure}[ht]
	\centering
	\includegraphics[width=\textwidth]{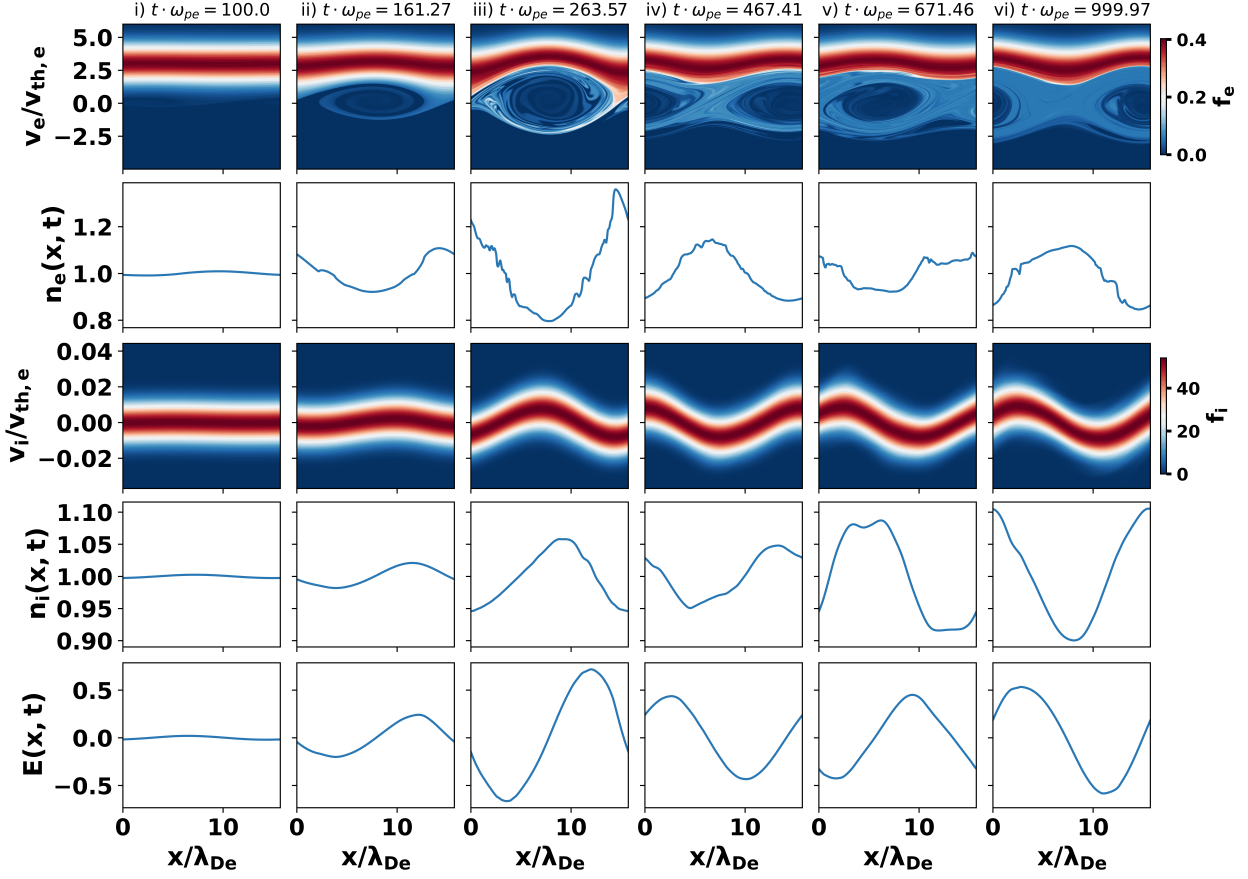}
	\caption{The Phase Space of Electron (first row) and Ion (third row) with their density profile (second and fourth row respectively) and the Electric field profile (fifth row) at various simulation time. The electron phase space hole coupled with IAW after the saturation is observed.}
	\label{hole_soliton}
\end{figure}
\begin{figure}[ht]
	\centering
	\includegraphics[width=0.8\textwidth]{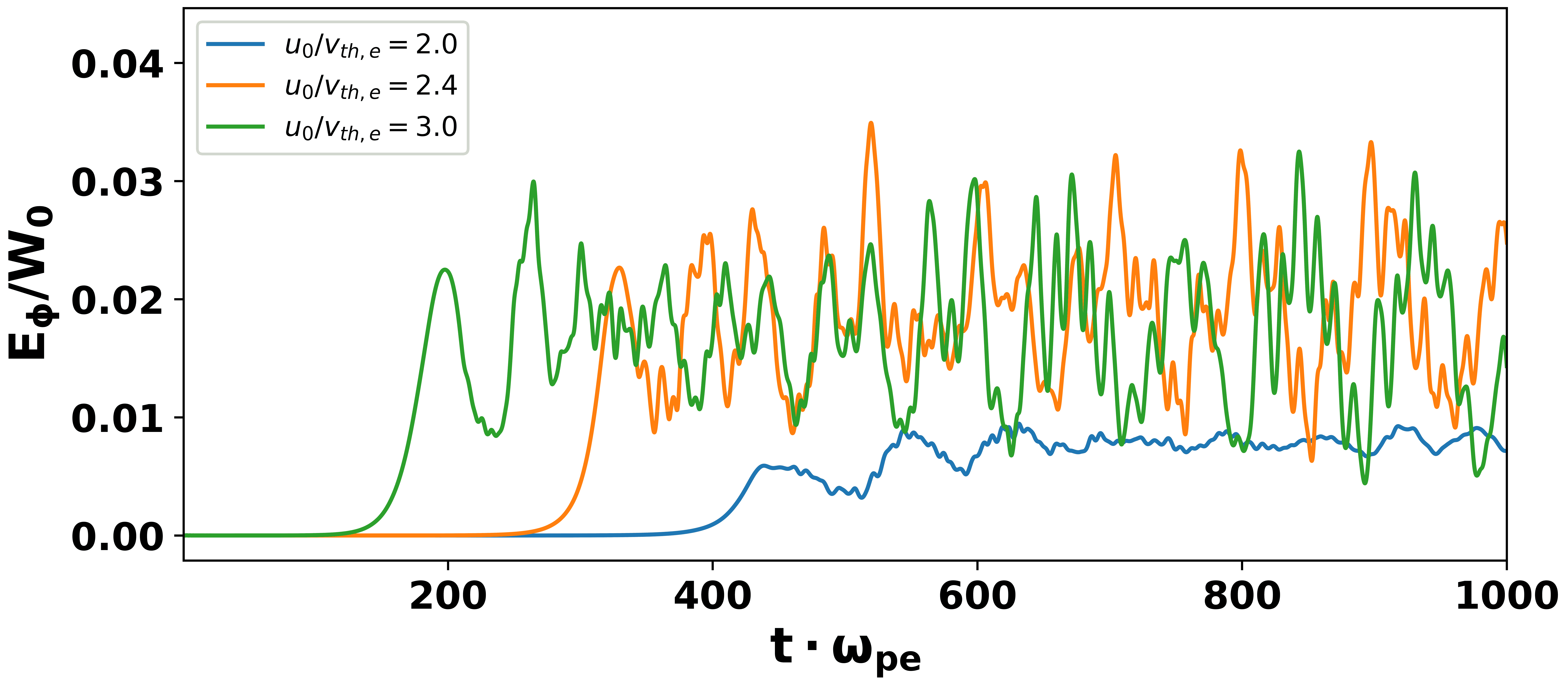}
	\caption{The Temporal Evolution of Ratio of Electrostatic Energy with Initial Kinetic Energy i.e., $E_{\phi}/\text{W}_{0}$. It is observed that the electrostatic energy increases during the linear growth regime, but no surge of potential energy at the time of saturation and the energy level after the saturation is much less than $0.1 \text{W}_{0}$.}
	\label{warm_potential_kinetic}
\end{figure}

Consider the ($k = 0.4$, $u_{0} = 3.0$) electron-perturbed simulation, the initial perturbation gets Landau damped as shown in Figure \ref{elec_fourier_pert} \& \ref{electric_mode_warm} for Fourier perturbation (given to the wave number $k_{perp} = 0.4$) and Figure \ref{white_noise_fourier_mode} for white noise perturbation; such damping have been observed in the past studies\cite{watt2002ion, petkaki2003anomalous, petkaki2006anomalous}. The higher harmonics are generated self consistently due to nonlinear dynamics of the system. In warm plasma limit, the Fourier modes' amplitude are orders apart before and after the saturation (refer Figure \ref{electric_mode_warm}) -- meaning $k_{min} = 0.4$ remains the dominant mode throughout the simulation. A consequence of this is that there no tearing of electron phase space hole as seen in cold plasma limit (see first row of Figure \ref{hole_soliton}). In the region of linear growth (first row - column ii in Figure \ref{hole_soliton}), non-linear features i.e., electron trapping are found to develop but the density steepening phenomenon is not observed. After the saturation i.e. $t\cdot\omega_{pe} > 200$ (first and fourth row - column iii, iv, v and vi in Figure \ref{hole_soliton}), electron hole is formed coupled with IAW(s). It is further observed that there is remnant streaming/beam electron\cite{hutchinson2024a} indicating the beam/streaming electron do not fully inject all the beam energy into the instability and thermalization of electron -- consequently the absence of energy spike at the time of saturation in the temporal evolution of the ratio of electrostatic energy and initial kinetic energy i.e., $E_{\phi}/\text{W}_{0}$ as shown in Figure \ref{warm_potential_kinetic}; the oscillatory behavior after the saturation as seen in cold plasma limit is also absent, instead a fluctuating electrostatic energy much below $0.1\text{W}_{0}$ is observed. 

In the quasi-linear study, Pavan \textit{et al.}\cite{pavan2011quasilinear} considered the initial electron velocity distribution function to be drifting Maxwellian. The time evolution of the electron velocity distribution is explicitly solved and the resulting velocity distribution is fed into a kinetic dispersion relation (similar to Eqn. \ref{PDFR}).  As the distribution function evolves, flattening develops around $v=0$ due to non-linear effects, leads to the decrease in the growth rate of the instability for all the modes. In another study performed by Tavassoli \textit{et al.}\cite{tavassoli2021role}, the kinetic dispersion relation is derived for a particular distorted electron Maxwellian velocity distribution which introduces correction to the existing dispersion relation (Eqn. \ref{PDFR}) where it was shown that the growth rate increases when such correction is taken into account -- the maximum growth rate doubles when 0.2\% density correction is applied to the initial electron velocity distribution. The flattening of the electron distribution function is due to non-linear dynamics i.e., electron trapping and such flattening increases with thermalization of electron. This would lead to higher and higher growth rate as the simulation transition from the linear to non-linear phase, but it is not so as constant growth rate is maintained during the linear phase of the instability.

The initial electron perturbation has the phase velocity, $v_{\phi} = 10.16$ for cold plasma and $v_{\phi} = 3.21$ for warm plasma, which is near the thermal bulk of the streaming electron in both cases. This initial perturbation interacts with the streaming electron by wave-particle interaction i.e., trapping/de-accelerating electrons causing the amplitude of the perturbation to increase. The heavier ions start to resposnd to the changes in electric field, thus generating IAWs. The IAWs behave as a quasi-stationary inhomogenous ion density background at electron time scale leading the excitation of higher harmonics by wave-wave interaction\cite{kaw1973quasiresonant, pandey2022coupling}.

In Kaw \textit{et al.}'s work\cite{kaw1973quasiresonant}, the ions are considered to be immobile with a sinusoidal density profile creating an inhomogeneous background. The electron perturbation interacts with the ion inhomogeneity leading to the generation of side-bands. The coupling parameter defined as $N^{2} = \epsilon/\gamma k_{i}^{2}$ where $\epsilon$ is the amplitude of ion inhomogeneity, $\gamma$ is the adiabatic constant of ion and $k_{i}$ is the inhomogeneity mode number, helps in determining how many side-bands will be generated. The energy of the primary mode cascades to the side-bands by mode-mode coupling; if a large number of side-bands are generated, the energy of the primary mode gets efficiently transfer to the side-bands. 
At the time where the higher harmonics get generated -- $t\cdot\gamma_{B} = 3.78$ for cold plasma and $t\cdot\gamma_{B} = 1.352$ for warm plasma (where $\gamma_{B}$ is the growth rate), the amplitude of ion inhomogeneity is $1\%$ and $0.2\%$ respectively giving the ratio of coupling parameter to be $N^{2}_{c}/N^{2}_{w} \approx 83$. As is clear, higher number of side-bands are generated in the cold plasma limit as compared to warm plasma limit, 
thus the amplitude of the higher harmonics become comparable to that of fundamental harmonic in the former while they are two orders apart in the latter (as shown in Figure \ref{electric_mode_cold} and \ref{electric_mode_warm} respectively).
Physically the density steepening is possible in cold plasma regime as there is no thermal pressure to oppose the density compression, but in the warm plasma regime the thermal pressure of the species oppose the density compression. Such a suppression in the steepening leads to formation of shallow potential explaining the absence in potential energy spike, which is the indication of the complete trapping of streaming/beam electron by the wave as well as the remnant of the streaming/beam electron. 
Additionally, the amplitude of the soliton decreases while its width increases with increase in ion temperature\cite{tappert1972improved, sakanaka1972formation, tagare1973effect}.


In summary, the results presented here show that the linear growth rate of the Buneman Instability deviate from the linearized fluid as well as kinetic dispersion relation at warm plasma limit. Such deviation in growth rate is not accounted for in the linearized kinetic model nor in the quasi-linear theory of the instability where velocity distortion is self-consistently developed. Furthermore, density steepening reduces as one transition from cold plasma to warm plasma limit due to the reduction in the generation of side-bands in warm plasma. Such phenomenon leads to formation of shallow potential which trap only a small portion of the streaming/beam electron, thus suppressing the complete transfer of energy from beam to the bulk plasma. In addition to the characteristic $M_{r}^{-1/3}$ scaling of the maximum growth rate, it is shown that the maximum growth rate is nearly independent of the temperature ratio ($T_{r}$).

\section{\label{conclusion}Conclusion}
The effects of thermal spread of the constituent species on Buneman Instability have been explored in present study using high resolution Vlasov-Poisson simulation. In previous studies of the instability, PIC simulation is employed where the particle are initialized with velocity distribution of negligible width at the beginning of the simulation. The simulation is performed with the MPI parallelized Vlasov-Poisson Solver (refer Appendix \ref{benchmarking} for benchmarking of the developed code). In addition to the numerical study, various theoretical model of plasma -- fluid and kinetic, have been considered for the derivation of linearized dispersion relation which helps in validating the simulation results. 

In the present work, the cold plasma limit results of the instability have been successfully reproduced using the Vlasov Solver. It is observed that the fluid dispersion relation (Eqn. \ref{WPDR}) generally overestimates the growth rate of the instability while the kinetic dispersion (Eqn. \ref{LKDR} and Eqn. \ref{PDFR}) provides accurate growth rates for small values of Doppler shift but tends to underestimate them at higher values in warm plasma limit. The overlapping of the two Kinetic Dispersion Solvers (shown in Figure \ref{growth_rate}) with regards to the value of growth rate indicates the accuracy of the developed Solvers. It has been shown that the growth rate of the instability is independent of which species is getting perturbed and the mode of perturbation, though the perturbation on ion species leads to faster saturation of the instability. The coupled electron hole-ion soliton is observed in the cold plasma limit, while in warm plasma limit the electron phase space hole is coupled with IAW(s) and also exists alongside streaming/beam electron, a remnant of the initial streaming/beam electron. It has been found that side-bands are excited due to the IAW(s) acting as a quasi-stationary inhomogenous ion background at electron time scale; the mode-mode coupling is more efficient in cold plasma as compared to warm plasma, thus steep potential can be formed in the cold plasma limit. Physically, the thermal pressure of the species oppose the density compression leading to the formation of a shallow potential structure, trapping/de-accelerating only a small portion of the streaming/beam electron. Furthermore, the dependence of the maximum growth rate on the mass ratio and the temperature ratio has been systematically examined. The results show that while the maximum growth rate exhibits the well known dependence on the mass ratio i.e., $M_{r}^{-1/3}$, it is essentially independent of the temperature ratio, $T_{r}$.

The finding of the present study is of importance to the understanding of astrophysical plasma where streaming of charge particles are prevalent and also laboratory plasma such as initial stage of current drive in tokamak, ion acoustic studies in linear devices. To the best of our knowledge, the existing linearized and quasilinear theory are not able to account for the deviation in the growth rate of the instability, thus it needs to be explored in more details.

\begin{acknowledgments}
	All the computational experiments are performed on Institute for Plasma Research (IPR) HPC Linux Cluster ANTYA. The authors would like to thank the Data Center staff at IPR.
\end{acknowledgments}

\appendix
\section{\label{decoupling}Thermal Effects on Buneman Instability's Growth Rate at Warm Plasma Fluid Limit.}
Consider the fluid equations,
\begin{equation}
	\frac{\partial n_{s}}{\partial t} + \frac{\partial }{\partial x}n_{s}u_{s}= 0
\end{equation}
\begin{equation}
	m_{s}n_{s} \left(\frac{\partial u_{s}}{\partial t} +u_{s} \frac{\partial}{\partial x}u_{s}\right)= q_{s}n_{s}E -\nabla p_{s}
\end{equation}
\begin{figure}[ht]
	\begin{subfigure}{0.9\textwidth}
		\includegraphics[width=\linewidth]{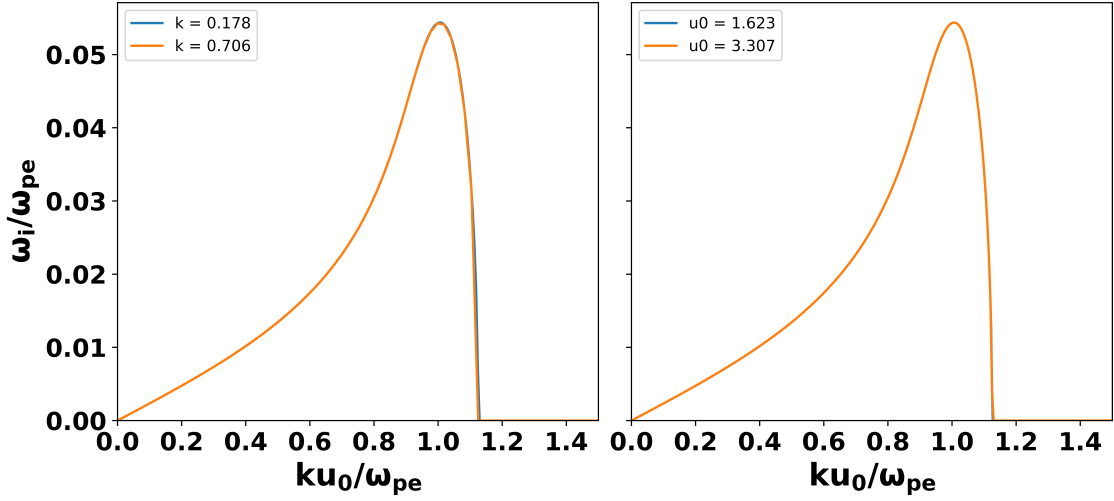}
		\caption{Cold Plasma Dispersion Relation. Parameter : $\gamma_{e} = \gamma_{i} = 0$ and $M_{r} = 1836$.}
		\label{fluid_cold}
	\end{subfigure}
	
	\begin{subfigure}{0.9\textwidth}
		\includegraphics[width=\linewidth]{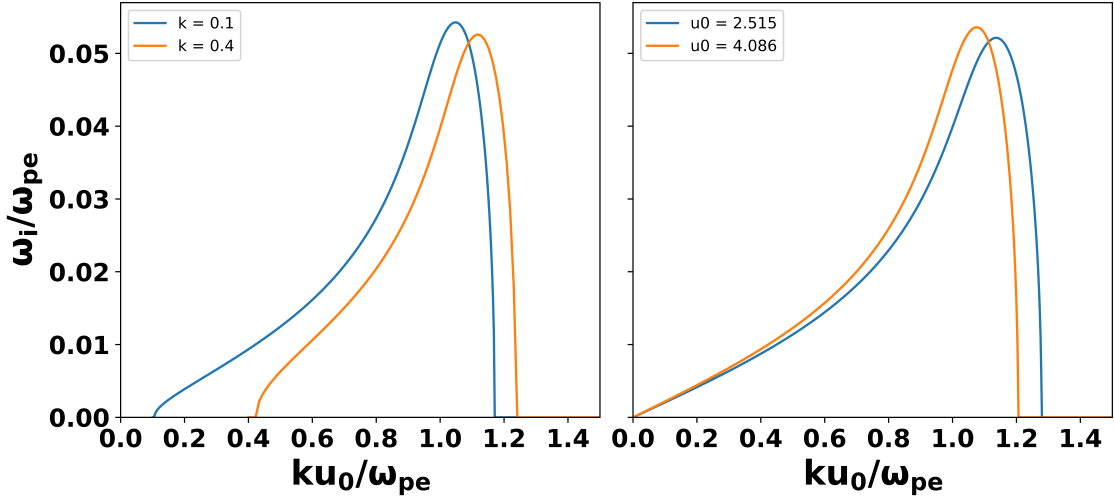}
		\caption{Warm Plasma Dispersion Relation. Parameter : $\gamma_{e} = \gamma_{i} = 1.0$, $M_{r} = 1836$ and $T_{r} = 0.1$.}
		\label{warm_fluid}
	\end{subfigure}
	\caption{Numerical Solution of the Fluid Dispersion Relations - Cold Plasma and Warm Plasma highlighting the thermal effect on growth rate.}
	\label{decouple_fig}
\end{figure}
where $s = i, e$. The above equation are linearized and coupled with the Poisson equation to get the warm plasma fluid Dispersion Relation (see Eqn. \ref{WPDR}). To solve it numerically, it is normalized with electron parameter to give,
\begin{equation}
	1  - \frac{1}{(\omega - u_{0}k)^{2}\displaystyle\bigg[1-\gamma_{e}\frac{k^{2}}{(\omega -u_{0}k)^{2}}\bigg]} -  \frac{1}{M_{r}\omega^{2}\displaystyle\bigg[1 - \gamma_{i}\frac{k^{2}}{\omega^{2}}\frac{T_{r}}{M_{r}} \bigg]}=0
	\label{Normalised_warm}
\end{equation}
where $\gamma_{e}$ and $\gamma_{i}$ is the adiabatic constant of electron and ion respectively. The above equation can be reduced to cold plasma fluid limit by setting $\gamma_{e} = \gamma_{i} = 0$. The concerned equation (Eqn. \ref{Normalised_warm}) is then converted into a quadratic polynomial
\begin{eqnarray}
	\omega^{4} - 2u_{0}k\,\omega^{3} 
	&+& \bigg[(u_{0}k)^{2} - \Big(\gamma_{e} - \gamma_{i}\frac{T_{r}}{M_{r}}\Big)k^{2} - \frac{1}{M_{r}} - 1\bigg]\omega^{2} \nonumber \\
	&+& \bigg[\frac{2u_{0}k}{M_{r}} + 2\gamma_{i}\,u_{0}k^{3}\frac{T_{r}}{M_{r}}\bigg]\omega 
	-\frac{(u_{0}k)^{2}}{M_{r}} 
	+ \gamma_{e}\gamma_{i}\frac{T_{r}}{M_{r}}k^{4} 
	+ \gamma_{i}\,k^{2}\frac{T_{r}}{M_{r}} 
	+ \gamma_{e}\frac{k^{2}}{M_{r}} = 0
\end{eqnarray}
After this, $\omega = \omega_{r} + i\omega_{i}$ is introduced into the polynomial to separate out the real and imaginary part, which are then solved with Newton-Raphson method.

The growth rates i.e., the imaginary roots ($\omega_{i}$) of the polynomial for $M_{r} = 1836$, $T_{r} = 0.1$ and $\gamma_{e} = \gamma_{i} = 1.0$ with streaming velocity $0 \leq u_{0}\leq 15$ and wave number $0 \leq k \leq 1.5$ are solved as stated above. It is observed that the growth rate in the cold plasma limit only depends on the Doppler shift; a symmetry exists between two cases -- fixing $k$ and varying $u_{0}$ or fixing  $u_{0}$ and varying $k$ (see Figure \ref{fluid_cold}). In the warm plasma fluid limit, the growth rates corresponding to different values of $k$ with varying $u_{0}$ differs from one another; such behavior is also observed when $u_{0}$ is fixed and $k$ is varied (see Figure \ref{warm_fluid}).

\section{\label{benchmarking}Benchmarking VPPM-MPI 1.0}
To upgrade the VPPM-OMP 1.0 (OpenMP)\cite{pandey2021landau, pandey2021trapped_a, pandey2021trapped_b, pandey2022coupling, pandey2024interaction} to Message Passing Interface (MPI) parallelization, the simulation domain  $(0, L) \times (-v_{s}^{max}, v_{s}^{max})$, where $s= i, e$ needs to be decomposed such that the computational load can be shared among the processes. Slab or 1D decomposition along the velocity domain has been implemented - such data structure has been termed Space-Major Data Structure i.e, for a range of velocity the concerned space information is available to the process. Ghost cell has been introduced to facilitate the velocity advection. In such a arrangement, a process only needs to communicate with its neighbor processes.
\begin{figure}[ht]
	\centering
	\begin{subfigure}{0.55\textwidth}
		\centering
		\includegraphics[width=\linewidth]{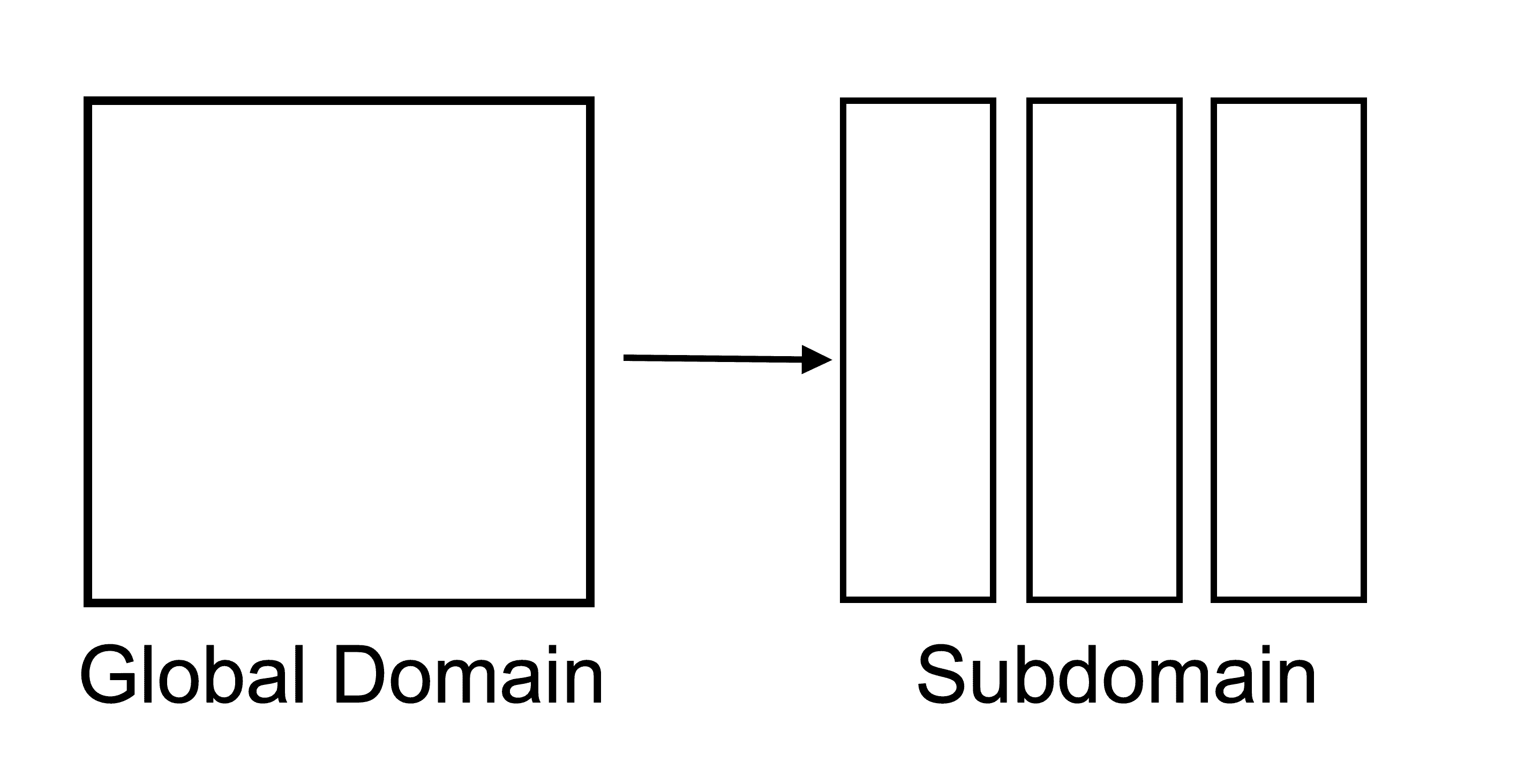}
		\caption{Slab or 1D Decomposition}
	\end{subfigure}
	\begin{subfigure}{0.4\textwidth}
		\includegraphics[width=\linewidth]{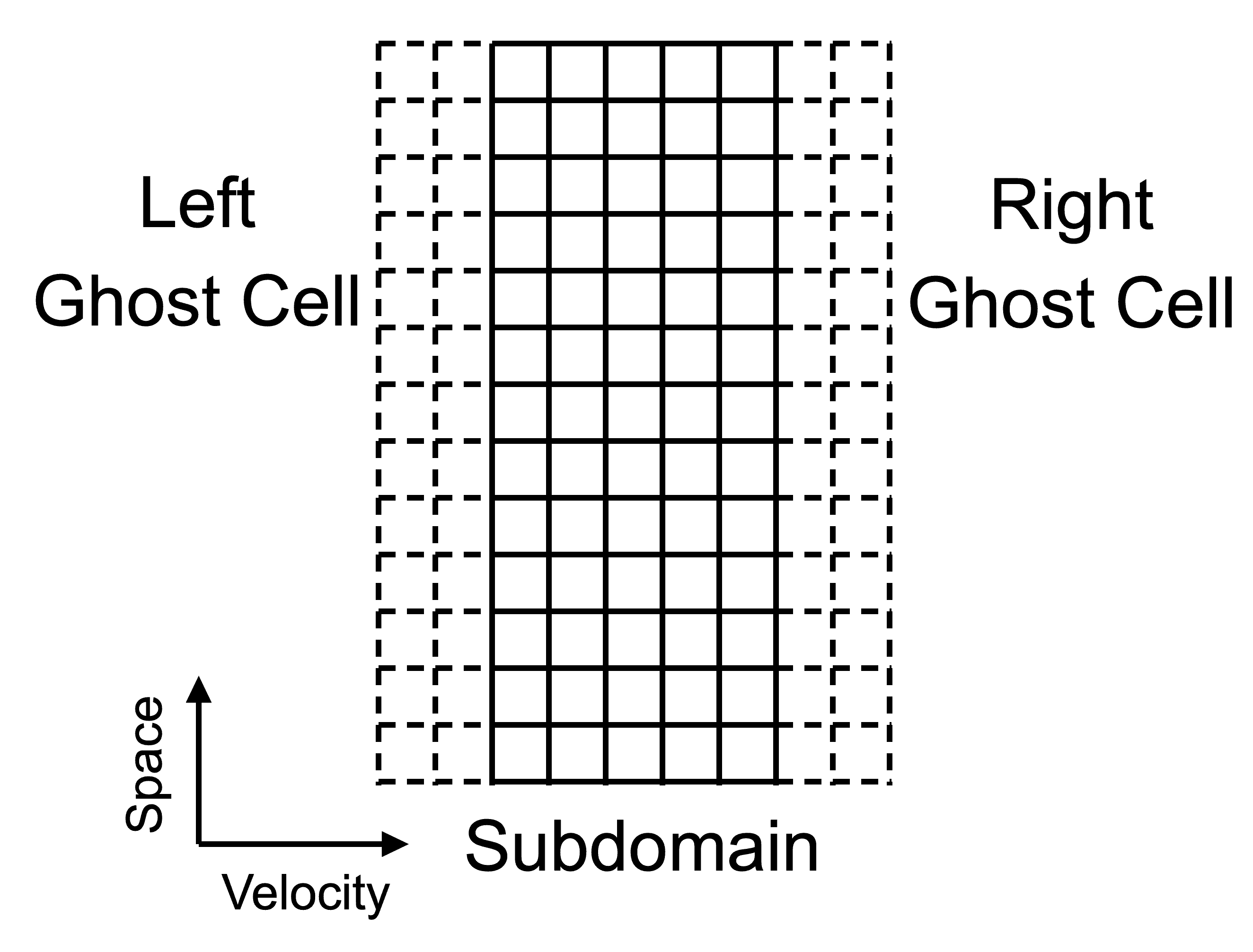}
		\caption{Schematic of Sub-Domain}
	\end{subfigure}
	\caption{Information on the Domain Decomposition and Sub-Domain.}
\end{figure}

Manfredi's non-linear Landau damping problem\cite{manfredi1997long} is solved to benchmark the newly developed VPPM-MPI 1.0 Code. Parameter of the run are as follows: $(N_{x}, N_{v}) = (4096, 4096)$, $\alpha = 0.05$, $T_{r} = 1.0$, $M_{r} = 1836$, $dt_{max} = 0.1$ and $k=0.4$. 
\begin{figure}[ht]
	\centering
	\includegraphics[width=0.8\textwidth]{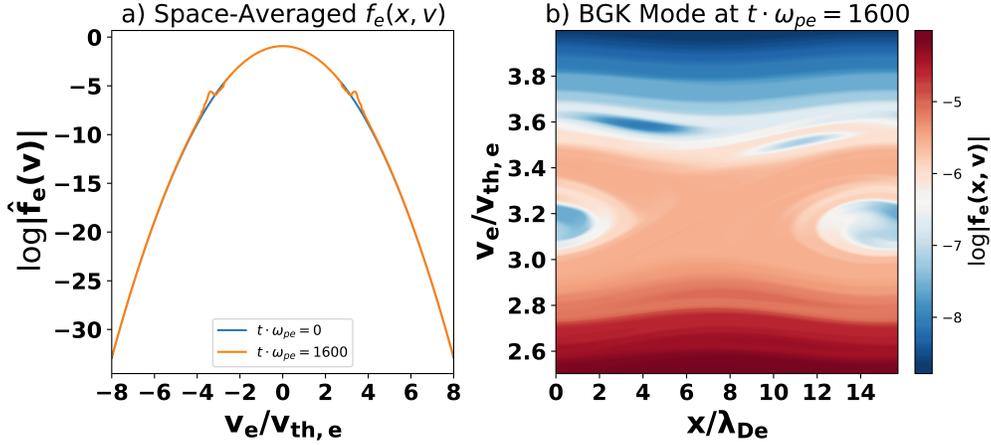}
	\caption{Electron Distribution Function: (a)  the space-averaged $f_{e}$ at $t\cdot\omega_{pe} = 0 \text{ and } 1600$ highlighting the distortion of $\hat{f}_{e}$ at phase velocity, $v_{\phi} = 3.21$ due to wave-particle interaction and (b) BGK Mode at $v_{\phi} = 3.21$ highlighting the dynamic of trapping and detrapping of particles.}
	\label{dist_func}
\end{figure}
\begin{figure}[ht]
	\centering
	\includegraphics[width=0.8\textwidth]{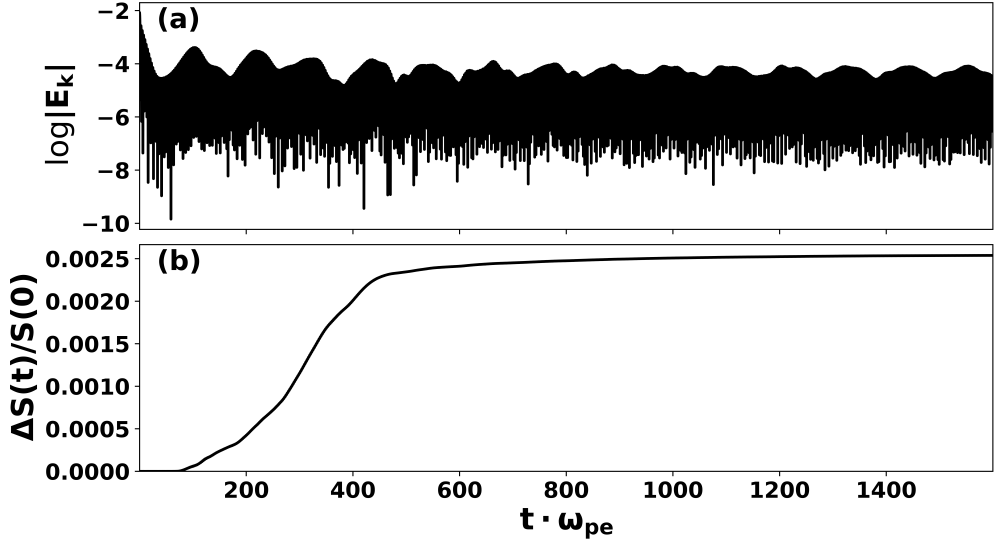}
	\caption{Temporal Evolution of (a) dominant Fourier Electric mode, $E_{k_{min} = 0.4}$ and (b) Relative Entropy, $\Delta S_{e}(t)/S_{e}(t=0)$.}
	\label{time_evo}
\end{figure}

The BGK mode is formed at the phase velocity, $v_{phi} \approx 3.2$ as shown in Figure \ref{dist_func}; the analytical value of $\omega  = 1.28506$ for $k=0.4$ i.e., $v_{\phi} = 1.28506/0.4 = 3.212$ which is in agreement from the numerical simulation. The phase structure is formed due to  the non-linear interaction between the wave and the particle i.e., trapping of electron in the potential. The time evolution of the dominant Fourier electric mode and the relative entropy is shown in Figure \ref{time_evo}. Initially, the wave gets Landau damped; as the system evolves, the wave-particle interaction sustains the wave. Both the temporal plots are in agreement with Manfredi's results.

\nocite{*}

\bibliography{apssamp}

\end{document}